\documentclass[usenatbib,useAMS,letterpaper]{mnras}
\usepackage{float}
\usepackage{graphicx}
\usepackage{fixltx2e}
\usepackage{epstopdf}
\usepackage{ textcomp }
\usepackage{multirow}
\usepackage{ctable}
\usepackage{url}
\usepackage{times}
\usepackage{amsmath}
\usepackage{amssymb}
\usepackage{xspace}
\usepackage{mathrsfs} 
\usepackage{tabularx}
\usepackage{fixltx2e}
\usepackage{color}
\usepackage{placeins}
\usepackage{comment}

\newcommand{\acknowledgments}{\begin{small}\section*{Acknowledgements}\end{small}}

\newcommand{\paperone}{Paper {\small I}}

\newcommand\sref[1]{\hyperref[#1]{\S~\ref*{#1}}}
\newcommand\fref[1]{\hyperref[#1]{Fig.~\ref*{#1}}}
\newcommand\Eqref[1]{equation~(\hyperref[#1]{\ref*{#1}})}
\newcommand\tref[1]{\hyperref[#1]{Table~\ref*{#1}}}
\newcommand\aref[1]{\hyperref[#1]{Appendix~\ref*{#1}}}

\newcommand{\mr}[1]{\multirow{2}{*}{#1}}
\newcommand{\mc}[1]{\multicolumn{1}{l}{#1}}

\title[What Types of Feedback Quench?]{Cosmic Rays or Turbulence can Suppress Cooling Flows (Where Thermal Heating or Momentum Injection Fail)}
\author[Su, Hopkins, Hayward et al.]{
\parbox[t]{\textwidth}{
Kung-Yi Su$^{1}$\thanks{E-mail: ksu@caltech.edu}, Philip F. Hopkins$^{1}$, Christopher C. Hayward$^{2,3}$,   Claude-Andr\'e Faucher-Gigu\`ere$^4$, Du\v san Kere\v s$^5$, Xiangcheng Ma$^{1,6}$,  Matthew E. Orr$^{1}$, T. K. Chan$^{5}$, Victor H. Robles$^{7}$
}
\vspace*{6pt} \\
$^1$TAPIR 350-17, California Institute of Technology, 1200 E. California Boulevard, Pasadena, CA 91125, USA\\
$^2$Center for Computational Astrophysics, Flatiron Institute, 162 Fifth Avenue, New York, NY 10010, USA\\
$^3$Harvard-Smithsonian Center for Astrophysics, 60 Garden Street, Cambridge, MA 02138, USA\\
$^4$Department of Physics and Astronomy and CIERA, Northwestern University, 2145 Sheridan Road, Evanston, IL 60208, USA\\
$^5$Department of Physics, Center for Astrophysics and Space Sciences, University of California at San Diego, 9500 Gilman Drive, La Jolla, CA 92093, USA\\
$^6$Department of Astronomy and Theoretical Astrophysics Center, University of California Berkeley, Berkeley, CA 94720, USA\\
$^7$Center for Cosmology, Department of Physics and Astronomy, University of California, Irvine, CA 92697, USA
}
\usepackage[all]{hypcap}

\begin{document}
\long\def\/*#1*/{}
\date{Submitted to MNRAS}

\pagerange{\pageref{firstpage}--\pageref{lastpage}} \pubyear{2018}

\maketitle

\label{firstpage}

\begin{abstract}
The quenching ``maintenance'' and ``cooling flow'' problems are important from the Milky Way through massive cluster elliptical galaxies. Previous work has shown that some source of energy beyond that from stars and pure magnetohydrodynamic processes is required, perhaps from AGN, but even the qualitative form of this energetic input remains uncertain. Different scenarios include thermal ``heating,'' direct wind or momentum injection, cosmic ray heating or pressure support, or turbulent ``stirring'' of the intra-cluster medium (ICM). We investigate these in $10^{12}-10^{14}\,{\rm M}_{\odot}$ halos using high-resolution non-cosmological simulations with the FIRE-2 (Feedback In Realistic Environments) stellar feedback model, including simplified toy energy-injection models, where we arbitrarily vary the strength, injection scale, and physical form of the energy. We explore which scenarios can quench without violating observational constraints on energetics or ICM gas. We show that turbulent stirring in the central $\sim100\,$kpc, or cosmic-ray injection, can both maintain a stable low-SFR halo for $>$Gyr timescales with modest energy input, by providing a non-thermal pressure which stably lowers the core density and cooling rates. In both cases, associated thermal-heating processes are negligible. Turbulent stirring preserves cool-core features while mixing condensed core gas into the hotter halo and is by far the most energy efficient model. Pure thermal heating or nuclear isotropic momentum injection require vastly larger energy, are less efficient in lower-mass halos, easily over-heat cores, and require fine-tuning to avoid driving unphysical temperature gradients or gas expulsion from the halo center.
\end{abstract}


\section{Introduction} \label{S:intro}
How to ``quench'' the {\it massive} galaxies and keep them ``red and dead'' over a large fraction of cosmic time, at stellar masses $\gtrsim 10^{11}\,{\rm M}_{\odot}$ (above $\sim L_{\ast}$ in the galaxy luminosity function), has been a major outstanding problem in galaxy formation for decades \citep[see e.g.][]{2003ApJS..149..289B,2003MNRAS.341...54K,2003MNRAS.343..871M,2004ApJ...600..681B,2005MNRAS.363....2K,2005ApJ...629..143B,2006MNRAS.368....2D,2009MNRAS.396.2332K,2010A&A...523A..13P,2012MNRAS.424..232W}. The major difficulty lies in the classic ``cooling flow'' problem --- X-ray observations have found significant radiative cooling in the hot gas of elliptical galaxies and clusters, indicating cooling times shorter than a Hubble time \citep{1994ApJ...436L..63F,2006PhR...427....1P}. However, comparing the inferred cooling flow (reaching up to $\sim 1000\,{\rm M}_\odot {\rm yr}^{-1}$ in clusters), neither sufficient cold gas  from H{\scriptsize I} and CO observations  \citep{2011ApJ...731...33M,2013ApJ...767..153W} nor sufficient star formation \citep{2001A&A...365L..87T,2008ApJ...681.1035O,2008ApJ...687..899R} have been found in galaxies. Simulations and semi-analytic models, which do not suppress cooling flow and simply allow gas to cool into the galactic core, typically predict over an order of magnitude higher star formation rates (SFRs) than observed \citep[for recent examples, see e.g., the weak/no feedback runs in][]{2007MNRAS.380..877S,2009MNRAS.398...53B,2015MNRAS.449.4105C,2015ApJ...811...73L}.

To compensate for the observed cooling, there must be some sort of heat source or pressure support. Moreover, the heat must still preserve the cool core structure in the majority of galaxies according to the observations \citep{1998MNRAS.298..416P,2009A&A...501..835M}.  One way to achieve this is to suppress the cooling flow and maintain a very-low-SFR stable cool-core (CC) cluster. Another possibility is that clusters undergo cool-core -- non-cool-core (NCC) cycles: a stronger episode of feedback overturns the cooling flows, resulting in a  non-cool-core cluster, which  gradually recovers to a cool-core cluster and start another episode of feedback. 

The various non-AGN solutions to the cooling flow problem  proposed in the literature generally belong to the former case, as they are mostly steady heating mechanisms.  These generally invoke physics that are un-ambiguously present, but play an uncertain role in quenching and/or the cooling flow problem, including: stellar feedback from shock-heated AGB winds \citep{2015ApJ...803...77C}, Type Ia supernovae (SNe) \citep[e.g.][and references therein]{2012MNRAS.420.3174S}, or SNe-injected cosmic rays (CRs) \citep{2017ApJ...834..208R,2017MNRAS.465.4500P,Buts18,Farb18,2018MNRAS.475..570J}; magnetic fields \citep{1990ApJ...348...73S,1996ARA&A..34..155B,2012MNRAS.422.2152B} and thermal conduction \citep{1981ApJ...247..464B,1983ApJ...267..547T,2002MNRAS.335L...7V,2002MNRAS.335L..71F,2003ApJ...582..162Z} in the circum-galactic medium (CGM) or intra-cluster medium (ICM); or ``morphological quenching'' via altering the galaxy morphology and gravitational stability properties \citep{2009ApJ...707..250M,2009ApJ...703..785D}. Although these processes can slightly suppress the star formation, and ``help'' suppress the cooling flows, most previous studies, including our own exhaustive survey studying each of these in simulations similar to those presented here \citep[][hereafter \paperone]{su:2018.stellar.fb.fails.to.solve.cooling.flow}, have shown that they do not fundamentally alter the classic cooling flow picture. In the end, the star formation is still cooling flow regulated, and the star formation rate is orders of magnitude too high.

Consequently, AGN feedback seem to be the most promising possible solution to the cooling flow problem, and there has been a tremendous amount of theoretical work on the topic (for recent studies see  \citealt[][]{2017ApJ...847..106L,2018arXiv180301444L,2017ApJ...837..149G,2017arXiv171004659W,2017MNRAS.468..751E,2017MNRAS.467.1449J,2017MNRAS.467.1478J,2018ApJ...856..115P,2018arXiv180303675Y,2018arXiv180506461M}; and see e.g., \citealt{1998A&A...331L...1S,1999MNRAS.308L..39F,2001ApJ...551..131C,2005ApJ...630..705H,2006ApJS..163....1H,2006MNRAS.365...11C,2007ARA&A..45..117M,2008MNRAS.384..251G,2009ApJ...699...89C,2010ApJ...722..642O,2012ApJ...754..125C,2013MNRAS.434.2209W,2013ApJ...779...10P} for earlier works). Observations show that the available energy budget can easily be comparable to the cooling rate, and un-ambiguous cases of AGN expelling gas from galaxies, injecting thermal energy via shocks or sound waves or photo-ionization and Compton heating, ``stirring'' the CGM and ICM, and creating ``bubbles'' of hot plasma with non-negligible relativistic components, are ubiquitous \citep[see e.g.,][for a detailed review]{2018arXiv180604680H}.

However, despite its plausibility and the extensive work above, the detailed physics of AGN feedback remains uncertain, as do the relevant ``input parameters.'' Unlike stellar feedback, where we have strong theoretical and observational constraints on  supernovae event  rates, energy inputs, metal yields, etc, AGN properties like energetics, kinetic luminosities, duty cycles, geometries,  and their dependence on the black hole mass and accretion are much less well-constrained. Besides, even with the same energy input rate, how and where the energy is coupled to the CGM and ICM  remain highly uncertain. 

Therefore, instead of jumping into a specific (potentially more realistic) AGN feedback model, in this study we ``take a step back'' and explore various idealized AGN ``toy models'' with energy injection in different forms (e.g., direct isotropic momentum injection, turbulent stirring, thermal heating, cosmic-ray injection), acting on different spatial scales, and with different energetics. Our goal is to answer the following simple questions: 
{\bf (a)} What form[s] of energy input (if any) can possibly quench a cooling flow, {\em without} generating un-realistic galaxy or halo properties in obvious disagreement with observations? For example, one could easily imagine scenarios which ``quench'' galaxies by simply expelling all the gas in the halo -- but this would violate the wealth of observations indicating massive halos retain most of the cosmological baryon fraction \citep[e.g.][]{2009ApJ...703..982G,2013ApJ...778...14G,2013MNRAS.429.3288S} (let alone more detailed constraints on density/temperature/entropy profiles).
{\bf (b)} If any form of energy injection is viable, over what (order-of-magnitude) spatial scales must it act? In other words, if the energy is primarily deposited around the galactic nucleus, does this yield behavior that is ``too explosive''? Does the injection have to be fine-tuned to occur where the cooling is occurring? 
{\bf (c)} Likewise, what are the required energetics, and are they reasonable compared to observational constraints and plausible accretion efficiencies of supermassive black holes in these systems? 
{\bf (d)} If a model quenches, what is the actual mechanism? For example, turbulent stirring could suppress cooling flows via heating through thermalized kinetic energy (viscous or shock-heating), or through providing non-thermal pressure which ``holds up'' the halo despite its cooling, or through bulk mixing of cold and hot gas.
{\bf (e)} Does the model quench by maintaining a low-SFR stable cool-core cluster or turning it into a non-cool-core cluster? If it is the latter case, how long (if ever) does it take to recover a cool core after the injection is turned off?

All of these questions have been studied to varying extent in the literature already (see references above). And we will argue below that our conclusions are largely consistent with this previous work. But this manuscript expands on these previous studies in at least three important ways. 
{\bf (a)} We attempt a broader and more comprehensive survey, across a variety of energy injection mechanisms, scales, and energetics, in different halo masses, using an otherwise identical set of physics and numerics, to enable fair comparisions. 
{\bf (b)} We aim to implement all of these in fully ``live,'' global simulations that self-consistently (and simultaneously) treat the entire halo and star-forming galactic disk. For such global simulations, our survey also reaches higher resolution compared to most previous work, allowing us to resolve more detailed sub-structure in the CGM and galactic disk.
{\bf (c)} We include explicit, detailed treatments of radiative cooling, the multi-phase ISM and CGM, star formation, and stellar feedback following the FIRE\footnote{FIRE project website: \href{http://fire.northwestern.edu}{\textit{http://fire.northwestern.edu}}} simulations \citep{2014MNRAS.445..581H,2015arXiv150103155M,2017arXiv170206148H}, in order to more robustly model both the gas dynamics and the response of galactic star formation rates to cooling flows.

In \sref{S:methods} we summarize the AGN toy models considered here, and describe our numerical simulations. Results are presented in \sref{S:results}. We then discuss the effects of each of these model in turn, in \sref{s:discussion}.

\section{Methodology} \label{S:methods}
Our simulations use {\sc GIZMO} \citep{2015MNRAS.450...53H}, \footnote{A public version of this code is available at \href{http://www.tapir.caltech.edu/~phopkins/Site/GIZMO.html}{\textit{http://www.tapir.caltech.edu/$\sim$phopkins/Site/GIZMO.html}}.} in its meshless finite mass (MFM) mode, which is a Lagrangian mesh-free Godunov method, capturing advantages of grid-based and  smoothed-particle hydrodynamics (SPH) methods. Numerical implementation details and extensive tests are presented in \cite{2015MNRAS.450...53H}.

Our default simulation uses the FIRE-2 implementation of the Feedback In Realistic Environments (FIRE) physical treatments of the ISM and stellar feedback, the details of which are given in \citet{2017arXiv170206148H,hopkins:sne.methods} along with extensive numerical tests.  Cooling is followed from $10-10^{10}$K, including the effects of photo-electric and photo-ionization heating, collisional, Compton, fine structure, recombination, atomic, and molecular cooling. Star formation is treated via a sink particle method, allowed only in  molecular, self-shielding, locally self-gravitating \citep{2013MNRAS.432.2647H} gas, above a density $n>100\,{\rm cm^{-3}}$. Star particles, once formed, are treated as a single stellar population with metallicity inherited from their parent gas particle at formation. All feedback rates (SNe and mass-loss rates, spectra, etc.) and strengths are IMF-averaged values calculated from {\small STARBURST99} \citep{1999ApJS..123....3L} with a \citet{2002Sci...295...82K} IMF. The feedback model includes: (1) Radiative feedback including photo-ionization and photo-electric heating, as well as single and multiple-scattering radiation pressure tracked in five bands  (ionizing, FUV, NUV, optical-NIR, IR). (2) Stellar particles continuously lose mass and inject mass, metals, energy, and momentum in the form of OB and AGB winds. (3) Type II and Ia SNe (including both prompt and delayed populations) happen stochastically according to the tabulated rate. Once they occur, the stellar particles lose mass and inject the appropriate mass, metal, momentum and energy to the surrounding gas.

\subsection{Initial Conditions}
\label{S:ic}

The initial conditions studied here are presented and described in detail in \paperone. Their properties are summarized in \tref{tab:ic}. In this paper, the bulk of our study will initially focus on the {\bf m14} halo, which has the most dramatic (massive) cooling flow (we will then consider the other halos in turn). The dark matter (DM) halo, bulge, black hole, and gas+stellar disk are initialized following  \cite{1999MNRAS.307..162S,2000MNRAS.312..859S}.
We assume a spherical, isotropic, \citet{1996ApJ...462..563N} profile DM halo; a \cite{1990ApJ...356..359H} profile stellar bulge; an exponential, rotation-supported disk of gas and stars ($10^{10}$ and $2\times10^{10} M_\odot$) initialized with Toomre $Q\approx1$; a BH with mass $1/300$ of the bulge mass \citep[e.g.][]{2004ApJ...604L..89H}; and an extended spherical, hydrostatic gas halo with a $\beta$-profile ($\beta=1/2$) and rotation at twice the net DM spin (so $\sim 10-15\%$ of the support against gravity comes from rotation, the rest thermal pressure resulting from the virial shock). The initial metallicity drops from solar ($Z=0.02$) to $Z=0.001$ with radius as $Z=0.02\,(0.05+0.95/(1+(r/20\,{\rm kpc})^{1.5}))$. For the runs with CR injection, initial magnetic fields are azimuthal with $|{\bf B}|=0.3\,\mu{\rm G}/(1+(r/20\,{\rm kpc})^{0.375})$ (extening throughout the ICM), and initial CR energy density is in equipartition with the local initial magnetic energy density. The ICs are run adiabatically (no cooling or star formation) to relax any initial transients before use. 

The ICs are designed to be similar to observed cool-core systems of similar mass wherever possible \citep[see e.g.][]{2012ApJ...748...11H,2013MNRAS.436.2879H,2013ApJ...775...89S,2015ApJ...805..104S}. Our {\bf m14} halo has initial cooling rate at $\sim 8\times10^{43}{\rm erg\,s}^{-1}$, with $\sim3 \times10^{43}{\rm erg\,s}^{-1}$ radiated in X-ray (0.5-7 kev). 

In {\bf m12} and {\bf m13} the mass resolution is constant; in {\bf m14} (given its much larger total mass but the need to ensure fixed physical mass resolution in e.g., the star-forming disk) the resolution here matches run ``MR-MRS'' in \paperone, adopting a radially-dependent super-Lagrangian refinement scheme. The target gas mass resolution is set to $= 3\times10^4\,{\rm M}_{\odot}$ inside $r < 10\,$kpc, and increases smoothly $\propto r$ outside outside this radius up to a maximum $=2\times 10^{6}\,{\rm M}_{\odot}$ at $\sim 300\,$kpc. Gas resolution elements are automatically merged or split appropriately if they move inward/outward, to maintain this mass resolution (to within a factor $=2$ tolerance) at all times. 
A resolution study is included in the appendix of \paperone. 

\begin{table*}
\begin{center}
 \caption{Properties of Initial Conditions for the Simulations/Halos Studied Here}
 \label{tab:ic}
 \begin{tabular*}{\textwidth}{@{\extracolsep{\fill}}lccccccccccccccc}
 \hline
\hline
&\multicolumn{2}{c}{\underline{Resolution}}&\multicolumn{3}{c}{\underline{DM halo}}&&\multicolumn{2}{c}{\underline{Stellar Bulge}}&\multicolumn{2}{c}{\underline{Stellar Disc}}&\multicolumn{2}{c}{\underline{Gas Disc}}&\multicolumn{2}{c}{\underline{Gas Halo}}  \\
$\,\,\,\,$Model  &$\epsilon_g$ &$m_g$        &$M_{\rm halo}$   &$r_{dh}$            &$V_{\rm Max}$    &$M_{\rm bar}$    &$M_b$ 
                 &$a$          &$M_d$        & $r_d$             &$M_{gd}$       &$r_{gd}$         &$M_{gh}$         &$r_{gh}$     \\
                 &(pc)         &(M$_\odot$)  &(M$_\odot$)      & (kpc)           &(km/s)           &(M$_\odot$)      &(M$_\odot$) 
                  &(kpc)        &(M$_\odot$)  &(kpc)            &(M$_\odot$)    &(kpc)            &(M$_\odot$)      &(kpc)\\
\hline
$\,\,\,\,$m12        &1       &8e3           &1.5e12           &25             &174              &2.2e11           &1.5e10   
                     &1.0       &5.0e10          &3.0                &5.0e9            &6.0                &1.5e11           &25        \\                    
$\,\,\,\,$m13        &3       &5e4           &1.0e13           &100              &240              &7.2e11           &1.0e11   
                     &2.8     &1.4e10        &2.8              &5.0e9           &2.8              &6.0e11              &10             \\
{\bf $\,\,\,\,$m14}  &{\bf1}  &{\bf3e4}      &{\bf8.5e13}      &{\bf 220}       &{\bf600}         &{\bf1.5e13}        &{\bf2.0e11}
                     &{\bf3.9}     &{\bf2.0e10}          &{\bf3.9}              &{\bf1e10}           &{\bf3.9}              &{\bf1.5e13}           &{\bf 22}             \\          
\hline 
\hline
\end{tabular*}
\end{center}
\begin{flushleft}
Parameters of the galaxy models studied here (\sref{S:ic}): 
(1) Model name. The number following `m' labels the approximate logarithmic halo mass. 
(2) $\epsilon_g$: Minimum gravitational force softening for gas (the softening for gas in all simulations is adaptive, and matched to the hydrodynamic resolution; here, we quote the minimum Plummer equivalent softening).
(3) $m_g$: Gas mass (resolution element). There is a resolution gradient for m14, so its $m_g$ is the mass of the highest resolution elements.
(4) $M_{\rm halo}$: Halo mass. 
(5) $r_{dh}$: NFW halo scale radius (the corresponding concentration of m12,m13,m14 is $c=12,\,6,\,5.5$).
(6) $V_{\rm max}$: Halo maximum circular velocity.
(7) $M_{\rm bar}$: Total baryonic mass. 
(8) $M_b$: Bulge mass.
(9) $a$: Bulge Hernquist-profile scale-length.
(10) $M_d$ : Stellar disc mass.
(11) $r_d$ : Stellar disc exponential scale-length.
(12) $M_{gd}$: Gas disc mass. 
(13) $r_{gd}$: Gas disc exponential scale-length.
(14) $M_{gh}$: Hydrostatic gas halo mass. 
(15) $r_{gh}$: Hydrostatic gas halo $\beta=1/2$ profile scale-length.
\end{flushleft}
\end{table*}

\subsection{Energy Injection Models Surveyed}
\label{S:physics}

\begin{figure}
\centering
\includegraphics[width=8.5cm]{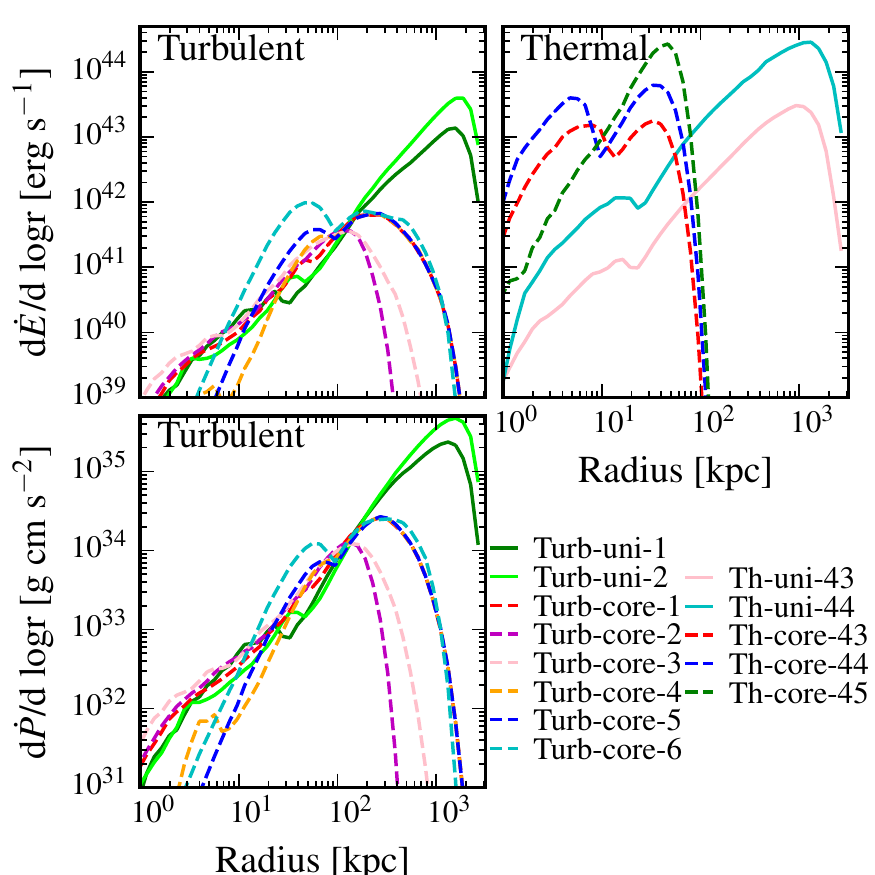}
\caption{Energy ({\em top}) or momentum ({\em bottom}) input rate per unit logarithmic galacto-centric radius $\log r$ (time-averaged over the last $100$\,Myr of each run), in a subset of our halo {\bf m14} runs. Runs labeled ``uni'' inject these quantities uniformly per unit mass over the whole halo, so large radii (containing most mass/volume) receive most of the injection. Runs labeled ``core'' have injection in a Guassian-like kernel, so most of the energy/momentum ends up around the kernel scale radius. Runs labeled ``BH'' inject everything in a kernel centered in the resolution elements immediately surrounding the BH ($\ll $\,kpc, hence not shown). 
}
\label{fig:energy_mom}
\end{figure}

The toy models we investigate include momentum injection (simulations prefixed ``Momm''), turbulent stirring (``Turb''), thermal input (``Th'') and CR input (``CR''). All the simulations are listed in \tref{tab:run}, which also tabulate the energy and momentum input within different ranges. The `Default' run includes only `FIRE-2' stellar feedback. The other runs have various AGN toy models implemented on top of `FIRE-2' stellar feedback. Only the runs with cosmic ray injection have magnetic fields.  The runs labeled as ``BH'' have energy (momentum) injected in the black hole neighborhood, while the ``core'' runs have a wider-distributed injection with the kernel functions listed in the last column of   \tref{tab:run}. The other runs labeled ``uni'' have uniform input per unit gas mass (so most of the energy is deposited at large radii). The detailed radial dependence of the energy and momentum input is shown in \fref{fig:energy_mom}.  The simulation duration is also listed in \tref{tab:run}. All runs are run to $2\,$Gyr, unless either the halo is completely ``blown out'' or completely un-affected. 

Although we will treat the energy/momentum injection rates as essentially arbitrary in our survey, for context it is worth noting that for a $\sim 10^{9}\,{\rm M}_{\odot}$ BH (about as massive as we expect in our {\bf m14} halo) the Eddington limit is $\sim 10^{47}\,{\rm erg\,s^{-1}}$. The associated photon momentum flux is $L/c \sim 4\times10^{36}\,{\rm g\,cm\,s^{-2}}$. For more typical low-luminosity AGN observed in massive galaxies, the energies associated with e.g.\ their jets reach $\sim10^{44}-10^{45}{\rm erg\,s}^{-1}$ (see \citealt{2012ARA&A..50..455F}).

\subsubsection{Thermal Input (``Pure Heating'')}
\label{S:thermal}

Any process that ultimately transfers some energy to gas thermal energy can be said to have a ``heating'' component. This can occur via radiative (photo-ionization, Compton), mechanical (shocked winds/jets, compression), viscous (damped sound waves or turbulence), cosmic ray (collisions, streaming instabilities), and other processes. Many models in the literature have invoked the idea that heating from AGN can effectively offset cooling and drive strong pressure-driven outflows, if roughly a few percent of the luminosity associated with near-Eddington phases can couple thermally \citep{2004cbhg.symp..374B,2005MNRAS.361..776S,2005Natur.433..604D,hopkins:qso.all,hopkins:bhfp.theory,hopkins:groups.qso,hopkins:red.galaxies,hopkins:twostage.feedback,2009ApJ...690..802J,2010ApJ...722..642O,2012MNRAS.425..605F,2014MNRAS.437.1456B,2013MNRAS.428.2885D,2017MNRAS.465.3291W,2018MNRAS.473.4077P,2018MNRAS.474.3673R,2018MNRAS.478.3100R}.

To mimic this in an intentionally idealized and simplified manner, we directly add (to the usual self-consistent heating and cooling routines) an analytic heating rate per unit mass $\dot{e}_{\rm inj}(r) = \dot{E}_{\rm tot}\,M_{0}^{-1}\,f(r)$, where $f(r)$ is a dimensionless spherically-symmetric kernel function (centered on the BH at the galaxy center) normalized to $M_{0}^{-1}\int \rho({\bf x})\,f(|{\bf x}|)\,d^{3}{\bf x} = 1$. We vary both $\dot{E}_{\rm tot}$ and $f(r)$ systematically, as shown in \tref{tab:run}. In runs labeled ``BH,'' $f(r)$ is a cubic spline with radius of compact support enclosing the nearest $\sim 96$ gas elements to the BH. In runs labeled ``core,'' $f(r)$ is a Gaussian ($\propto \exp{[-(r/r_{0})^{2}]}$) with dispersion approximately equal to the $\beta$-profile scale-length (which is also approximately the critical cooling radius). And in runs labeled ``uni,'' $f(r)$ is constant out to approximately the virial radius. 
In the ``BH'' cases, $f(r)$ is updated at each time step, while in the ``core'' and ``uni’’ cases, $f(r)$ is set at the beginning of the runs, and kept constant.\footnote{
This causes the evolution of energy input, especially in the more explosive runs, since the density profiles also evolve.}



\subsubsection{Momentum Input}
\label{S:kinetic}

Again many processes can transfer momentum/kinetic energy to gas, including radiation pressure, mechanical feedback from AGN winds and jets, and ``PdV'' work from cosmic ray pressure gradients. Again many models have invoked kinetic feedback to suppress cooling flows and SFRs in massive halos \citep{2011MNRAS.411..349G,2012ApJ...754..125C,2015MNRAS.449.4105C,2015ApJ...811...73L,2018arXiv180506461M} and many have argued it specifically provides a better match to observational constraints and is more efficient compared to ``pure heating'' models, especially in the context of ``maintenance'' or ``radio mode'' feedback \citep{2004cbhg.symp..374B,2012ARA&A..50..455F,2013MNRAS.428.2885D,2014MNRAS.437.1456B,2017ApJ...841..133M,2017MNRAS.465.3291W,2018MNRAS.473.4077P,2018arXiv180506461M}.

Since we will distinguish ``random'' or ``non-oriented'' driving below, we use this term to refer specifically to models with kinetic feedback oriented strictly radially away from the BH. Moreover because the coupling in the models above is primarily local (and we are not interested for this model in e.g.\ the case of CRs or hot thermally-pressurized gas driving outflows on large scales, since these should be resolved in our ``Thermal Input'' and ``Cosmic Ray'' runs), we will primarily focus on just the ``BH'' (local-kernel) models in this case. In that case a constant momentum flux $\dot{P}$ (directed radially away from the BH) is injected in a similar kernel-weighted fashion among neighboring gas around the BH (as for thermal energy), but with the kernel weights proportional to the solid angle subtended by each gas element (as seen by the BH).\footnote{We emphasize that while this is launched at the BH, it is not a jet model.  The scaling with solid angle simply ensures that momentum is launched uniformly in all directions. }

\subsubsection{Turbulent Driving or ``Stirring''}
\label{S:turb_driving}

Rather than simply ``pushing outwards,'' a variety of processes can instead transfer energy to kinetic energy of bulk quasi-random motion, what we call ``turbulent stirring.'' AGN bubbles may generate turbulence through Rayleigh-Taylor (RT) and Richtmyer-Meshkov (RM) instabilities \citep{ 2006PhFl...18h5101D,2008ApJ...686..927S,2009MNRAS.398..548B}; jets (precessing or not) can drive turbulence through changing bulk motion or secondary instabilities \citep[e.g.][]{2014ApJ...789...54L,2016ApJ...818..181Y,2017MNRAS.472.4707B,2018arXiv180506461M} with driving scale $\sim 100\,$kpc \citep{2016ApJ...817..110Z,2018PASJ...70....9H}; and non-AGN processes like halo mergers \citep[e.g.][]{1993ApJ...407L..53R,1997ApJS..109..307R,1999LNP...530..106N,2001ApJ...561..621R,2009MNRAS.395..180M,2011ApJ...726...17P,2011A&A...529A..17V}, sloshing of cold fronts \citep[e.g.][]{2004ApJ...612L...9F,2013ApJ...762...78Z,2018ApJ...853..180Z}, and winds from satellites can do likewise \citep{2017MNRAS.470.4698A}. Studies have argued turbulence could suppress cooling flows by providing direct pressure support to gas \citep{2012MNRAS.419L..29P}, or heating the gas ``directly'' via viscous dissipation \citep{2014MNRAS.443..687B,2014Natur.515...85Z}, or mixing cold structures back into hot gas in a thermally-unstable medium and so efficiently re-distributing heat \citep[e.g.][]{2003ApJ...596L.139K,2006ApJ...645...83V,2010ApJ...712L.194P,2010ApJ...713.1332R,2014MNRAS.443..687B}.


We represent ``turbulent stirring'' by driving turbulence directly following the ``turbulent box'' simulations in \citet{2012MNRAS.423.2558B}. Turbulence is driven in Fourier space as an Ornstein-Uhlenbeck process \citep[see][] {2009A&A...494..127S,2010A&A...512A..81F,2010MNRAS.406.1659P} with characteristic driving wavelength ($\lambda=2\pi/k$) set to $1/2$ of the halo scale radius (experimenting with this, compared to the kernel or total energy, makes little difference to our conclusions). The compressive part of the acceleration is projected out via a Helmholtz decomposition in Fourier space so that the driving is purely incompressible (solenoidal). After Fourier-transforming back to real space, the stirring is applied as a continuous acceleration ${\bf a}({\bf x})$ to each element; at this stage, we apply the desired kernel function ${\bf a}({\bf x})\rightarrow {\bf a}({\bf x})\,f(r)\,V_{0}^{-1}$ (with $V_{0}^{-1}\int f(|{\bf x}|)\,d^{3}{\bf x}=1$).  In runs labeled ``uni,'' $f(r)$ is constant out to approximately the virial radius. In the runs labeled ``core,'' $f(r)$ is either a Gaussuan function or an exponential function as shown in \tref{tab:run}. The energy and momentum input rates labeled in \tref{tab:run} are calculated through $\dot{E}\sim \int dm\, \max(|{\bf a}({\bf x})|)|{\bf v}|$ and $\dot{P}\sim \int dm\, \max(|{\bf a}({\bf x})|)$, which estimate the upperbounds. \footnote{Although the acceleration of the gas (as a function of space) is constant in time, the density profiles change. Therefore, the total energy input rates also vary as a function of time.}

\subsubsection{Cosmic Ray Injection}
\label{S:cosmicray}

CRs arise generically from processes that result in fast shocks, so could come from shocked winds or outflows, but are particularly associated with relativistic jets from AGN (where they can make up the bulk of the jet energy; \citealt{2006PhRvD..74d3005B,2017ApJ...844...13R}) and hot, relativistic plasma-filled ``bubbles'' or ``cavities'' (perhaps inflated by jets in the first place) around AGN. Different authors have argued that they could help suppress cooling flows via providing additional pressure support to gas, driving pressure-driven outflows in the galaxy or CGM, or via heating the CGM/ICM directly via collisional (hadronic \&\ Coulomb) and streaming-instability losses \citep{2008MNRAS.384..251G,2010ApJ...720..652S,2011A&A...527A..99E,2011ApJ...738..182F,2013MNRAS.434.2209W,2013MNRAS.432.1434F,2017ApJ...834..208R,2017ApJ...844...13R,2013ApJ...779...10P,2017MNRAS.465.4500P,2017MNRAS.467.1449J,2017MNRAS.467.1478J,2018MNRAS.475..570J}.

We treat this analogous to our ``thermal heating'' runs -- simply injecting cosmic ray energy at some fixed rate within a kernel. The CR physics and numerical implementation are described in detail in \citet{chan:2018.cosmicray.fire.gammaray}. Briefly, this treats CRs including streaming (at the local Alfv\'en speed, with the appropriate streaming loss term, which thermalizes, following \citealt{Uhlig2012}, but with $v_{\rm st}=v_A$), diffusion (with a fixed diffusivity $\kappa_{\rm cr}$), adiabatic energy exchange with the gas and cosmic ray pressure in the gas equation of motion, and hadronic and Coulomb losses (following \citealt{2008MNRAS.384..251G}). We follow a single energy bin (i.e.\ GeV CRs, which dominate the pressure), treated in the ultra-relativistic limit. Streaming and diffusion are fully-anisotropic along magnetic field lines. In \citet{chan:2018.cosmicray.fire.gammaray}, we show that matching observed $\gamma$-ray luminosities, in simulations with the physics above requires $\kappa_{\rm cr}\sim 10^{29}\,{\rm cm^{2}\,s^{-1}}$, in good agreement with detailed CR transport models that include an extended gaseous halo around the Galaxy  \citep[see e.g.][]{1998ApJ...509..212S,2010ApJ...722L..58S,2011ApJ...729..106T}, so we adopt this as our fiducial value.\footnote{We caveat that we do not account for the possibility of different diffusion coefficient in different environments.} In practice, because of the large diffusivity, the CR energy density rapidly converges to the same quasi-equilibrium profile regardless of the shape of the injection kernel, so long as the injection scale is not extremely large ($\lesssim100\,$kpc), so we simplify by focusing on the ``BH'' kernel choice and keeping the injection isotropic.\footnote{We also note that, in the runs including CR heating, CRs from SNe are not included, so we have a clean test on the black hole CR injection. 
We showed in \paperone\ that CRs from SNe contribute negligibly to quenching, and we note below that the total energy injection from SNe is a factor $\sim 10^{2}-10^{4}$ below the analytically-input CR energy injection rate.}

\setlength{\tabcolsep}{4pt}
\begin{table*}
\begin{center}
 \caption{Physics variations (run at highest resolution) in our halo-{\bf m14} survey}
 \label{tab:run}
\resizebox{18cm}{!}{%
 \begin{tabular}{ccc|cccc|cccc|cc}
 \hline
\hline
Model           &Summary&$\Delta T$ &$\dot{E}_{tot}$ &$\dot{E}_{r<30}$ &$\dot{E}_{r<100}$ &$\dot{E}_{r>100}$ &$\dot{P}_{tot}$ &$\dot{P}_{r<30}$ &$\dot{P}_{r<100}$ &$\dot{P}_{r>100}$ &kernel\\
                &&(Gyr)      &\multicolumn{4}{c|}{(erg s$^{-1}$)}                                              &\multicolumn{4}{c|}{(g cm s$^{-2}$)}                                  &(r in kpc)\\
\hline
Default         &-&2.0     & -            & -             & -             &-             & -             & -             &-             & -             &- \\                                    
\hline
Momm-BH-34      & minor&0.4     & 4.7-7.6 e40     & -               & -               & -               & 1.2e34          & -               & -               & -               &BH neighbour\\                       
Momm-BH-35      & minor &2.0    & 1.8-3.6 e42     & -               & -               & -               & 1.2e35          & -               & -               & -               &BH neighbour\\                               
Momm-BH-36      &explosive&0.3     & 2.0-0.07 e44    & -               & -               & -               & 3.6e36          & -               & -               & -               &BH neighbour\\                       
\hline
Turb-uni-1      &minor&2.0     & 3.6-7.7 e42     & 5.2-17 e39      & 4.6-5.7 e40     & 3.5-7.6 e42     & 1.6-1.6 e35     & 3.3-5.1 e32     & 3.0-1.7 e33     & 1.6-1.6 e35     &Uniform\\                                 
Turb-uni-2      &$L_X\downarrow$&2.0     & 6.5-20 e42      & 9.5-11 e39      & 8.4-5.8 e40     & 6.4-20 e42      & 2.9-2.9 e35     & 6.0-3.4 e32     & 5.5-1.5 e33     & 2.9-2.9 e35     &Uniform\\                                 
Turb-core-1     &moderate&2.0     & 5.0-5.6 e41     & 8.6-15 e39      & 6.2-8.5 e40     & 4.4-4.8 e41     & 2.5-2.0 e34     & 5.5-4.5 e32     & 4.1-2.3 e33     & 2.1-1.8 e34     &$a\sim\exp(- r/200)\,\,\,\,\,\,\,\,$\\    
Turb-core-2     &moderate&2.0     & 1.6-2.4 e41     & 9.1-18 e39      & 6.7-11 e40      & 9.8-14 e40      & 9.8-7.8 e33     & 5.9-5.4 e32     & 4.5-2.9 e33     & 5.3-5.0 e33     &$a\sim \exp(-(r/140)^2)$\\                
Turb-core-3     &moderate&2.0     & 2.1-2.9 e41     & 1.6-2.1 e40     & 8.1-12 e40      & 1.3-1.6 e41     & 1.2-0.9 e34     & 9.8-6.1e32      & 5.4-3.1e33      & 6.6-6.2 e33     &$a\sim2\exp(- r/ 80)\,\,\,\,\,\,$\\       
\mr{\bf Turb-core-4}&\mr{\bf quenched}&\mr{\bf 2.0}&\mr{\bf 5.7-5.9 e41} &\mr{\bf 2.7-0.5 e40} &\mr{\bf 1.4-1.1 e41} &\mr{\bf 4.4-4.8 e41} &\mr{\bf 3.1-2.0 e34} &\mr{\bf 1.7-0.2 e33} &\mr{\bf 9.2-2.4 e33} &\mr{\bf 2.1-1.8 e34} &\mc{$\mathbf{a_{r<100}\sim3\exp(-(r/ 79)^2)}$}\\
&&&&&&&&&&                                                                                                                                                                &\mc{$\mathbf{a_{r>100}\sim \exp(- r/200)}$}\\   
\mr{Turb-core-5}&quenched&\mr{2.0}&\mr{6.6-6.6 e41} &\mr{5.2-1.7 e40} &\mr{2.2-1.8 e41} &\mr{4.4-4.8 e41} &\mr{3.6-2.1 e34} &\mr{3.3-0.3 e33} &\mr{1.5-0.3 e34} &\mr{2.1-1.7 e34} &\mc{$a_{r<100}\sim6\exp(-(r/ 66)^2)$}\\
&(NCC)&&&&&&&&&                                                                                                                                                                &\mc{$a_{r>100}\sim \exp(- r/200)$}\\      
\mr{Turb-core-6}&\mr{Mach$\uparrow$}&\mr{0.7}&\mr{9.9-11 e41}  &\mr{1.7-0.7 e41} &\mr{5.5-0.5 e41} &\mr{4.4-5.7 e41} &\mr{5.9-2.6 e34} &\mr{1.1-0.06 e34}&\mr{3.7-0.6 e34} &\mr{2.1-2.0 e34} &\mc{$a_{r<100}\sim20\exp(-(r/ 54)^2)$}\\
&&&&&&&&&&                                                                                                                                                                &\mc{$a_{r>100}\sim \exp(- r/200)$}\\
\hline
Th-uni-43          &minor&1.4      & 2.1e43          & 4.2-8.1 e40     & 3.9-4.0e41      & 2.0e43          & -               & -               & -               & -               &Uniform\\                         
Th-uni-44          &minor&2.0      & 2.1e44          & 4.2-7.3 e41     & 3.9-3.4 e42     & 2.0e44          & -               & -               & -               & -               &Uniform\\                                 
Th-core-43     &minor&2.0      & 2.0-1.8 e43     & 1.2-1.2e43      & 2.0-1.8e43      & 6.8-7.2 e38     & -               & -               & -               & -               &$\dot{E}\propto\exp(-(r/30)^2)$\\         
{\bf Th-core-44}     &{\bf quenched}&{\bf 2.0}      & {\bf 2.0-0.5 e44}     &{\bf 1.2-0.3 e44}     & {\bf 2.0-0.5 e44}     & {\bf 6.8-6.4 e39}     & -               & -               & -               & -               &$\mathbf{\dot{E}\propto\exp(-(r/30)^2)}$\\         
Th-core-45     &explosive&1.0      & 1.9-0.1 e45     & 1.2-0.02 e45    & 1.9-0.1 e45     & 6.8-5.2 e40     & -               & -               & -               & -               &$\dot{E}\propto\exp(-(r/30)^2)$\\ 
Th-BH-43       &minor&2.0      & 2.1e43          & -               & -               & -               & -               & -               & -               & -               &BH neighbour\\                               
Th-BH-44       &explosive&1.2      & 2.1e44          & -               & -               & -               & -               & -               & -               & -               &BH neighbour\\                       
Th-BH-45       &explosive&0.4      & 2.1e45          & -               & -               & -               & -               & -               & -               & -               &BH neighbour\\                       
\hline
CR-BH-42       &minor&2.0      & 2.1e42          & -               & -               & -               & -               & -               & -               & -               &BH neighbour\\                               
{\bf CR-BH-43}       &{\bf quenched}&{\bf 2.0}      &{\bf 2.1e43}          & -               & -               & -               & -               & -               & -               & -               &{\bf BH neighbour}\\                               
CR-BH-44       &explosive&0.3      & 2.1e44          & -               & -               & -               & -               & -               & -               & -               &BH neighbour\\                       
\hline 
\hline
\end{tabular}
}
\end{center}
\begin{flushleft}
This is a partial list of simulations studied here: each was run using halo {\bf m14}, systematically varying the energy injection mechanism, scale, and energetics, at our highest resolution (a broader low-resolution parameter survey, and our survey of halos {\bf m12} and {\bf m13}, are not included here). Columns list: 
(1) Model name. Models labeled ``Momm,'' ``Turb,'' ``Th,'' and ``CR'' correspond to (radial) momentum injection, turbulent injection or ``stirring,'' thermal energy injection (``heating''), and cosmic ray (CR) injection, respectively. Models labeled ``uni,'' ``core,'' ``BH'' adopt different kernels (see Fig.~\ref{fig:energy_mom}). 
(2) Summary of the results. Minor, moderate, and quenched correspond respectively to a SFR of $\gtrsim10$, $\sim1-10$, and $\lesssim 1 M_\odot\,{\rm yr}^{-1}$. The runs labeled otherwise are quenched while having a major drawback (as labeled).
(3) $\Delta T$: Simulation duration. All runs are run to $2\,$Gyr, unless either the halo is completely ``blown out'' or completely un-affected. 
(4) $\dot{E}_{tot}$, $\dot{E}_{r<30}$, $\dot{E}_{r<100}$, and $\dot{E}_{r>100}$ tabulate the total energy input of the corresponding spherical region (with the two values corresponding to the beginning and end of the run). The energy input of ``Momm'' and ``Turb'' runs is the energy used to accelerate gas (e.g.\ difference in kinetic energy) in each timestep. 
(5) $\dot{P}_{tot}$, $\dot{P}_{r<30}$, $\dot{P}_{r<100}$, and $\dot{P}_{r>100}$ tabulate the momentum input in the corresponding region. 
(6) kernel: the form of the injection kernel. 
\end{flushleft}
\end{table*}
\setlength{\tabcolsep}{6pt}

\section{Results in Our Massive Halo (m14) Survey} \label{S:results}
As will be shown in the following subsections, `Th-core-44' ($\dot{E}_{\rm th}\sim \dot{E}_{\rm cool}$), `Turb-core-4' ($\dot{E}_{\rm turb}<1\% \dot{E}_{\rm cool}$), and `CR-BH-43' ($\dot{E}_{\rm CR}\sim 10\% \dot{E}_{\rm cool}$) are the more successful runs in the corresponding toy model scenario. We, therefore, highlight these runs in the subsequent plots, while tuning down the contrast of the ``explosive'' runs.
\subsection{Star formation history} \label{S:sfr}

\begin{figure*}
\centering
 \includegraphics[width=16cm]{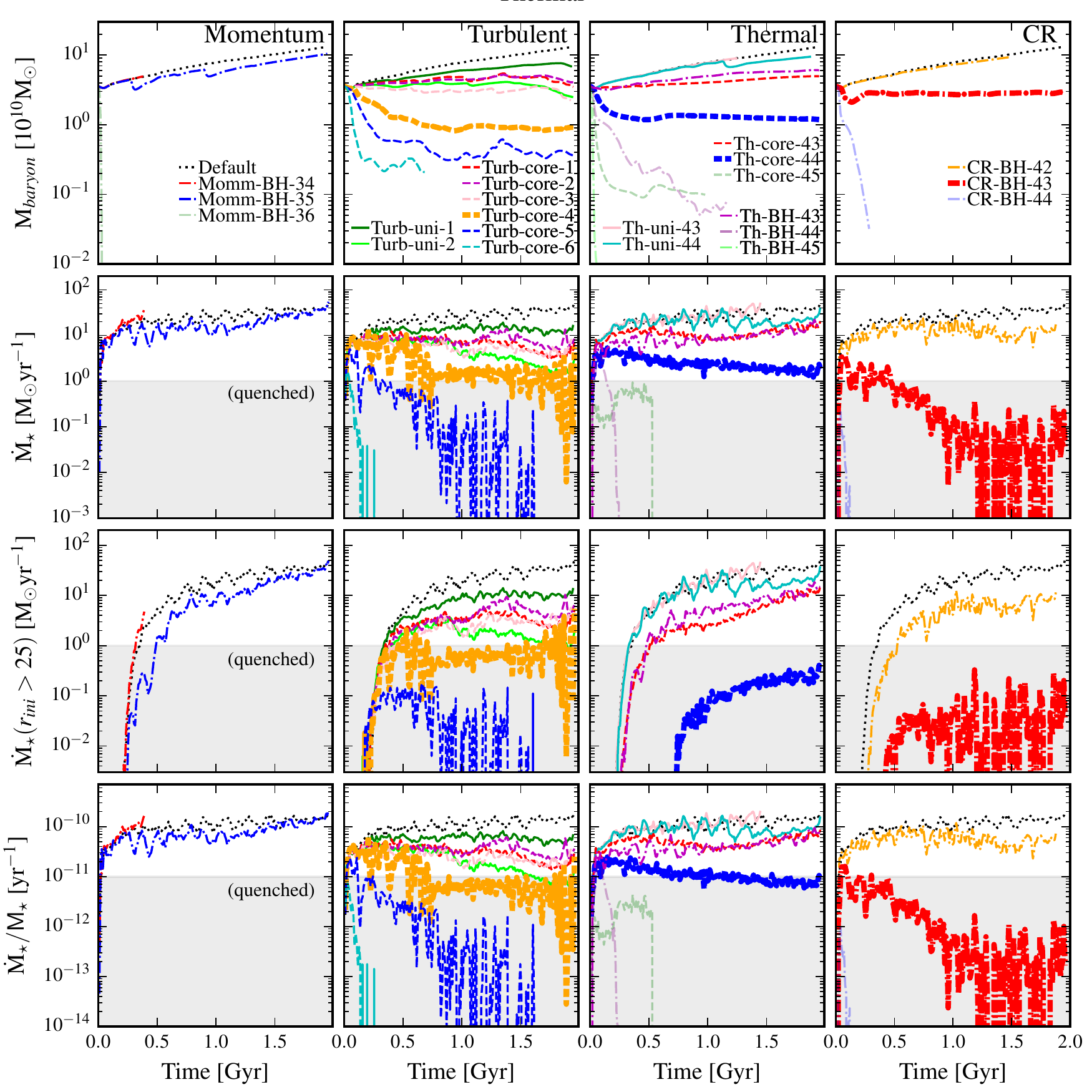}
\caption{{\em Top:} Baryonic mass within 30\,kpc (excluding pre-existing stars from the ICs), as a function of time, in the halo-{\bf m14} runs from Table~\ref{tab:run}. This is a proxy for the net amount of cooling-flow inflow.
{\em  Second:} SFRs averaged in 10\,Myr intervals. 
{\em  Third:} SFRs specifically from gas which was at $r>25\,$kpc in the ICs (gas which comes in with the cooling flow). 
{\em  Bottom:} Specific SFRs.
The shaded regions indicate the SFR or specific SFR that we define as quenched.
For each, we compare runs with momentum injection, turbulent stirring, thermal heating, and cosmic ray injection (columns, as labeled). 
Momentum injection below $\dot{P} \lesssim  10^{35} {\rm g\,cm\, s}^{-2}$ does not suppress cooling flows, while $\dot{P} \gtrsim 3\times10^{36} {\rm g\,cm\, s}^{-2}$ almost immediately ejects all the gas in the halo. 
Uniform thermal heating has little effect on SF (most of the energy is ``wasted'' at large-$r$), while nuclear (``BH'') injection transitions sharply between doing nothing (the heat is radiated away) and driving a Sedov-Taylor explosion that evacuates the halo around $\dot{E} \sim 3\times 10^{43}\,{\rm erg\,s^{-1}}$. Heating with a semi-extended $\sim 30\,$kpc kernel can suppress SF without explosive ejection for $\dot{E}$ carefully chosen around $\dot{E}\sim 10^{44} {\rm erg\,s}^{-1}$. 
Turbulent stirring more efficiently suppresses SF: when the driving $\dot{E}$ within $<100\,$kpc reaches $\gtrsim 10^{41} {\rm erg\, s}^{-1}$, the core baryonic mass begins to fall, and by $2\times$ this SF is eventually completely quenched. 
CR energy input at $10^{43} {\rm erg\,s}^{-1}$ can maintain a low SFR and semi-stable core baryonic mass even if the energy is deposited in the nucleus. 
}
\label{fig:sfr}
\end{figure*}

The first row of \fref{fig:sfr} plots the baryonic mass (as a function of time) within $30$ kpc ($M^{30\,{\rm kpc}}_{\rm baryon}$) excluding the pre-existing stars, which characterizes the cooling flow rates. The second, third and bottom rows show SFRs, SFRs from gas initially sitting outside 25 kpc (SFRs supplied by the cooling flows) and specific star formation rates (sSFRs), averaged in rolling $10\,$Myr bins.
Momentum injection below $\sim 10^{35} {\rm g\,cm\, s}^{-2}\sim 0.03\,L_{\rm Edd}/c$ does not suppress the cooling flow or star formation by much, while an injection above $3\times10^{36} {\rm g\,cm\, s}^{-2}\sim L_{\rm Edd}/c$ blows everything away within 50 Myr leaving almost no gas within 70 kpc. 

With a lower momentum flux ($1-2\times 10^{34} {\rm g\,cm\,s}^{-2}$; non-radial), turbulent stirring can significantly suppress the cooling flows and star formation. When the turbulent energy input within 100 kpc reaches $1.1- 1.4\times 10^{41} {\rm erg\,s}^{-1}$ (`Turb-core-4'), the core baryonic mass is suppressed by a factor of 3-10. For turbulent energy input rates above $\sim 2 \times 10^{41} {\rm erg\,s}^{-1}$ (`Turb-core-5,6'), the SF is eventually completely quenched.

Uniform thermal heating has little effect on the SFRs and cooling flows even if input rate reaches $\sim 10^{44} {\rm erg\,s}^{-1}$. 
Black hole thermal injection, on the other hand, undergoes a  sharp transition from having little effect to completely quenching the galaxy by blowing everything away (through a Sedov-Taylor explosion), between injection rates $10^{43}$ to $10^{44} {\rm erg\,s}^{-1}$ ($10^{-4} - 10^{-3} L_{\rm Edd}$). 
The transition is milder if the energy is smoothly injected within a Gaussian kernel of 30 kpc, in which case a stable core baryonic mass and low star formation rate can be maintain by a heating rate of  $\sim 10^{44}{\rm erg\,s}^{-1}$ (`Th-Core-44').  However, with a similar cooling flow rate (e.g., `Turb-core-1' and `Th-core-43'),  turbulent stirring suppresses SFR more efficiently (with a lower energy input rate) than core thermal heating.

Unlike thermal heating, CR energy input  can maintain a semi-stable core baryonic mass and suppressed SFR even if all the energy is deposited in the vicinity of the black hole.  The SFR and cooling flows are significantly suppressed by an energy input of $10^{43} {\rm erg\,s}^{-1}$, less than the rate required for a thermal heating run  with  Gaussian kernel to quench. However, when the CR input reaches $10^{44} {\rm erg\,s}^{-1}$, the resulting dramatic suppression of core baryonic mass becomes similar to what is caused by the ``explosive'' BH-kernel thermal heating.

\subsection{The resulting halo properties} \label{S:temp}

\subsubsection{Temperature, density, and entropy}

\begin{figure*}
\centering
\includegraphics[width=16cm]{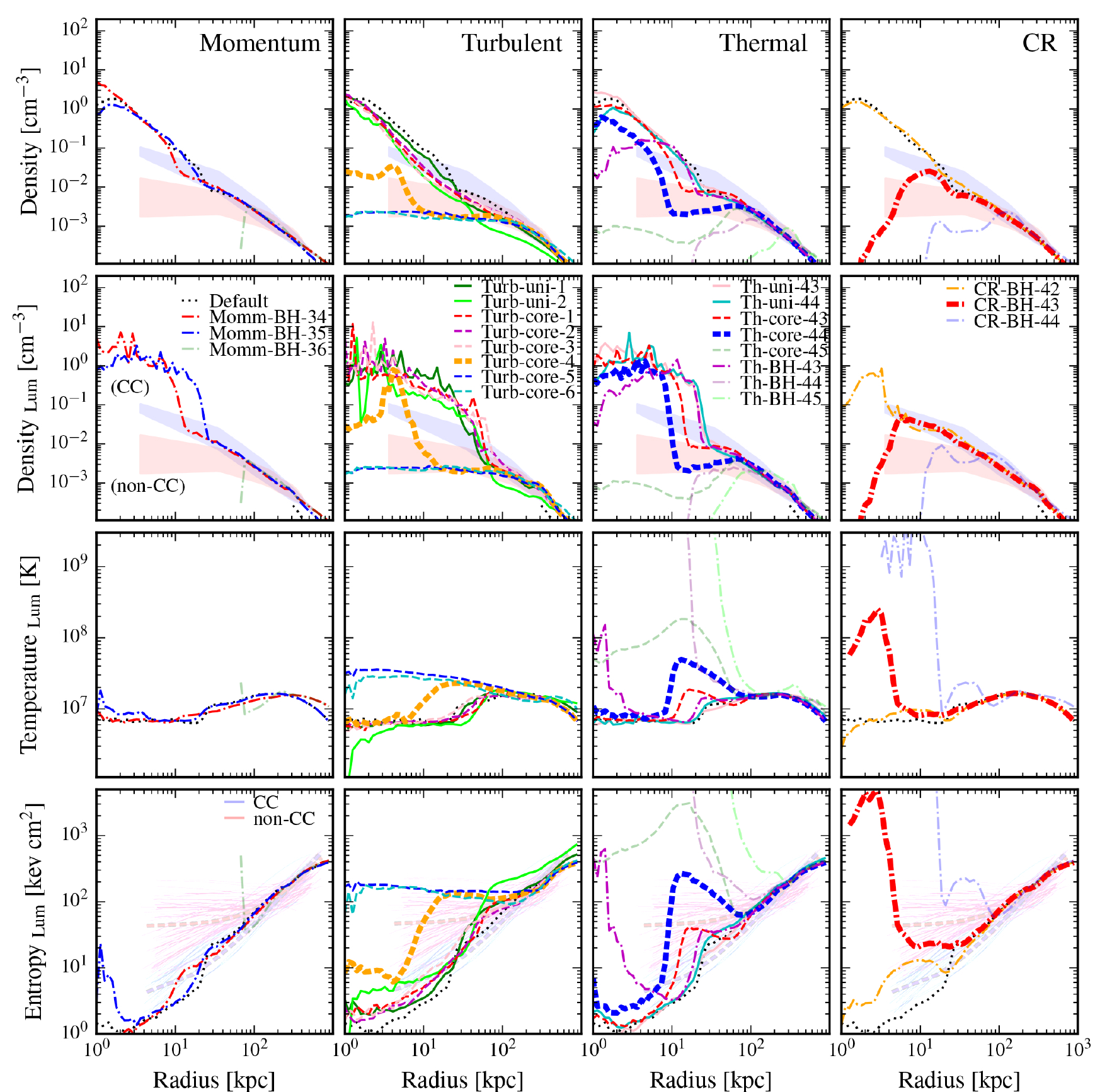}
\caption{Density ({\em top}), X-ray cooling luminosity-weighted density ({\em second}),  luminosity-weighted temperature ({\em third}), and luminosity-weighted entropy ({\em bottom}) versus radius averaged over the last $\sim 100\,$Myr in the {\bf m14} runs from \fref{fig:sfr}. 
The shaded regions in the second row and the lightened curves in the bottom row indicate the observational density and entropy profiles (scaled) for cool-core (blue) and non-core-core (red) clusters \citep{2013ApJ...774...23M}. 
At sufficiently low injection rates, all models do little (as expected). 
In ``Momentum,'' ``Thermal,'' and ``CR'' injection, we see that when the injection is nuclear (``BH'') and large enough, explosive behavior results (expelling nearly all gas within $\sim 30-100\,$kpc, and leaving what remains very hot), in stark contrast with observations.
Quasi-stable intermediate cases do exist, for turbulent stirring and CR injection in particular. 
Among the turbulent runs with suppressed SF, most preserve the initial cool-core features (though they do suppress the density, heat up, and flatten the entropy profile in the core), though ``core-5/6'' resemble non-cool-core clusters (but do not ``explode''); uniform turbulent driving suppresses densities even at $\gg 100\,$kpc as well.  The `Th-core-44' run, which has non-explosively suppressed SFR, broadly resembles non-cool-core clusters, but its negative temperature gradient is in tension with observations. The `CR-core-43' run, which also has non-explosively suppressed SFR, resembles cool-core clusters.
}
\label{fig:temp}
\end{figure*}

\fref{fig:temp} shows the average density, and luminosity-weighted density, temperature, and entropy as a function of radius averaged over the last $100\,$Myr of the runs.
The shaded regions in the second row  indicate the observational density profiles (scaled) for cool-core (blue) and non-core-core (red) clusters \citep{2013ApJ...774...23M}.\footnote{We use the panel for $z<0.1$ in  Fig.9 of \cite{2013ApJ...774...23M} and assume $\rho_{\rm crit}\sim9.2\times 10^{30} {\rm g \,cm}^{-3}$ and $r_{500}=650$ kpc (our m14 initial condition).} 
The lightened curves in the bottom row  indicate the observational entropy  profiles for cool-core (blue) and non-core-core (red) clusters \citep{2013ApJ...774...23M}.\footnote{The halos in \cite{2013ApJ...774...23M} have a mass range of $\sim2 \times 10^{14} < M_{500} < 20 \times 10^{14} M_\odot/h_{70}$.  We use their Fig.2 and scale the average entropy at $r=700$ kpc to 500 kev cm$^2$ given our halo is smaller (cooler). }

Momentum injection does not affect the resulting halo profiles with an input rate less than $\sim 10^{35} {\rm g\, cm\, s}^{-2}$, while it blows everything away when the input rate reaches $L_{\rm Edd}/c\sim 3\times10^{36} {\rm g\,cm\,s}^{-2}$.

With a lower momentum input at $\sim 10^{34}{\rm g\,cm\,s}^{-2}$, turbulent stirring can much more efficiently suppress the core density. When the core ($r<100$ kpc) energy input reaches $1.1-1.4 \times 10^{43} {\rm erg\,s}^{-1}$  (`Turb-core-4'), the density suppression becomes more significant. When it reaches $2 \times 10^{43} {\rm erg\,s}^{-1}$  (`Turb-core-5,6'), the core gas is eventually completely heated up, and the entropy profile is flattened.  If turbulent stirring is not suppressed at large radii, the density beyond $100\,$kpc is also suppressed by almost a factor of 10, i.e.\ the halo begins to expel/lose a significant amount of gas. 
Among the runs with significantly suppressed SFRs, the density and entropy profiles of `Turb-core-5' and `Turb-core-6'  end up resembling those observed in non-cool-core clusters \citep[compare][]{2006MNRAS.372.1496S,2009MNRAS.395..764S,2010A&A...513A..37H,2013ApJ...774...23M}, while `Turb-core-4' lives between cool-core and non-core. The other turbulent stirring runs with moderately suppressed  SFRs preserve the cool-core features, although their densities in the core regions are slightly higher than observational values.

The effects of thermal heating on the halo properties strongly correlate with the kernel size of the injection. When concentrated in the black hole neighborhood, $\sim10^{43} {\rm erg\,s}^{-1}$ is sufficient to significantly suppress the density within 5 kpc and heat up the gas up to $10^8$ K. Thermal injection rates in the BH neighborhood $\gtrsim10^{44} {\rm erg\,s}^{-1}$ blow out everything within 10\,kpc, heat gas to $\gtrsim 10^{10}$K, and produce a negative temperature slope out to $>100$\,kpc. If the injection is smoothed over a Gaussian kernel of 30 kpc, then the core density is not suppressed until the total energy input reaches $\gtrsim10^{44} {\rm erg\,s}^{-1}$ (when the energy input is comparable to the cooling).  Although milder, a negative temperature gradient extending from 10 to 100 kpc is still hard to avoid in that case.

CR injection can significantly suppress the core density with $\dot{E}\gtrsim 10^{43} {\rm erg\,s}^{-1}$, and produces an extended region with significant hot gas. If the input exceeds $10^{44} {\rm erg\,s}^{-1}$, the injection becomes explosive on large scales (similar to high-$\dot{E}$ BH-kernel thermal injection). Except for the explosive one, runs with CR injection have density and entropy profiles resembling those observed in cool-core clusters. The gas of the most successful CR injection run (`CR-BH-43') within $\sim7$ kpc is dominated by the hot gas from stellar mass loss (and/or gas heated by CR) and is less constrained by the observations. 

\begin{figure*}
\centering
\includegraphics[width=16cm]{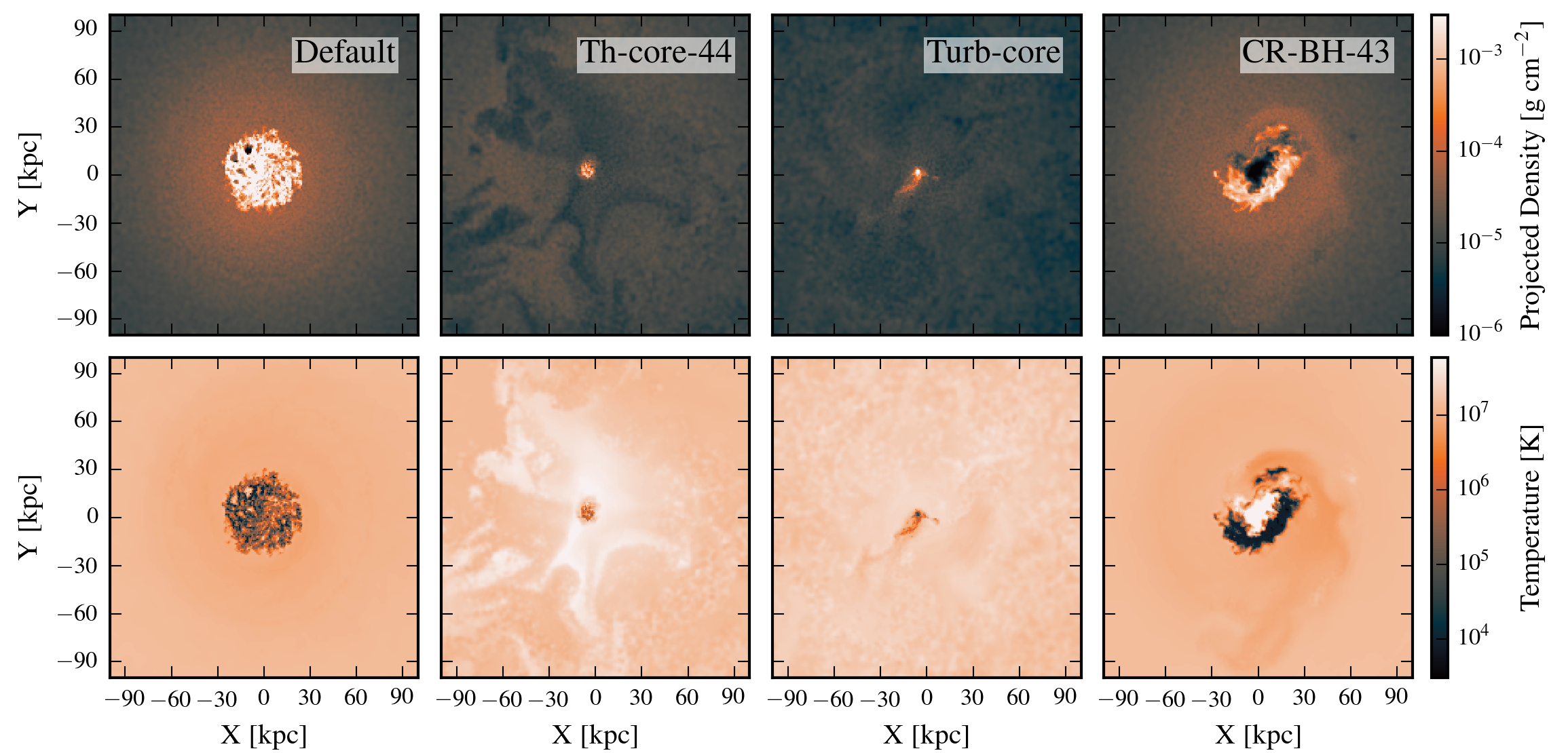}
\caption{The face-on projected density and average temperature (between $\Delta Z =\pm 1$ kpc) of the more successful runs (`Th-core-44', `Turb-core-4' and `CR-BH-43'). Consistent with the density and temperature profiles in  \fref{fig:temp},`Th-core-44' and `Turb-core-4' have suppressed density up to a few 10 kpc, while `CR-BH-43' has suppressed density only within 10 kpc. Thermal heating and CR injection both lead to a heated region, but the heated region in the former case extends to a larger radius. }
\label{fig:morph}
\end{figure*}

The face-on projected density and average temperature (between $\Delta Z =\pm 1$ kpc) of the more successful runs (`Th-core-44', `Turb-core-4' and `CR-BH-43') are shown in \fref{fig:morph}. Consistent with the aforementioned density and temperature profiles, `Th-core-44' and `Turb-core-4' have suppressed density up to a few 10 kpc, while `CR-BH-43' has suppressed density only within 10 kpc. Thermal heating and CR injection both lead to a heated region, but the heated region in the former case extends to a larger radius.


\subsubsection{X-ray luminosities} \label{S:luminosity}
\begin{figure}
\centering
\includegraphics[width=8.5cm]{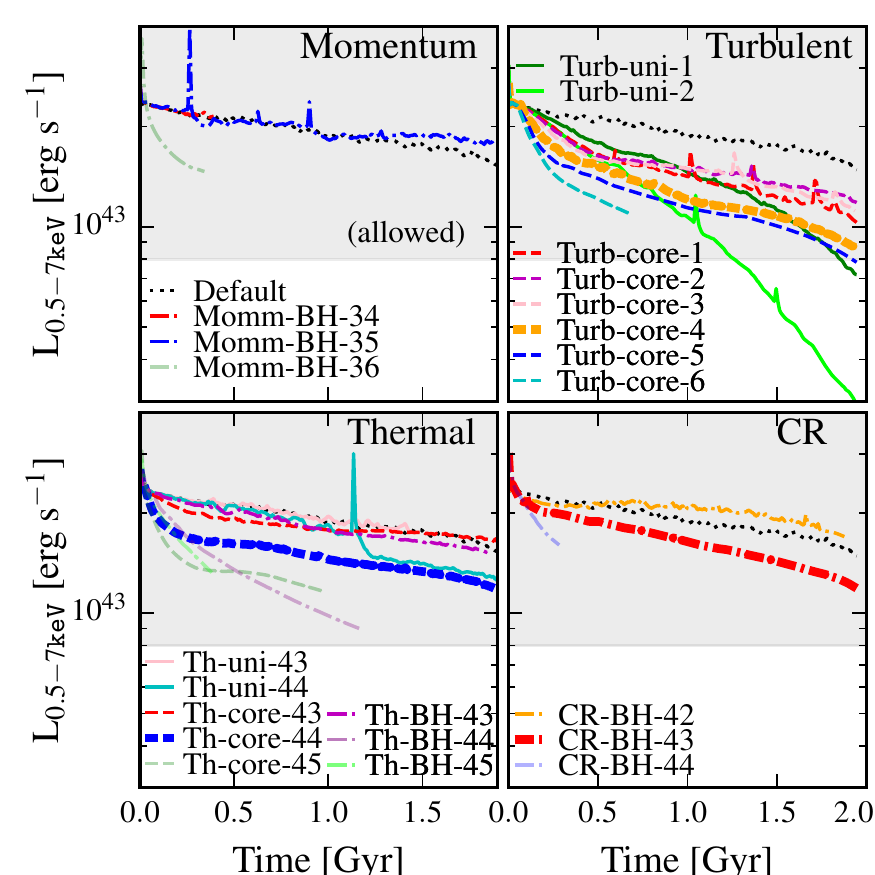}
\caption{X-ray cooling luminosity $L_{X}$, integrated from $0.5-7$\,keV, in the runs in Fig.~\ref{fig:temp}. The shaded regions indicate the observed  X-ray luminosities in \citet{2002ApJ...567..716R} and \citet{2006ApJ...648..956S} for halos with $m_{\rm halo}\sim 0.7-1.5 \times 10^{14} M_\odot$. Runs which explosively eject core gas (e.g.\ `Momm-BH-36', `Th-BH-44,45' and `CR-BH-44') strongly suppress $L_{X}$. Uniform turbulent stirring (`Turb-uni’) also suppresses $L_{X}$ strongly by ejecting gas (at larger radii). But other runs with suppressed SF (`Turb-core-X',`Th-Core-44', `CR-BH-43') have only factor $\sim 1.5-3$ lower $L_{X}$. This is because a large portion of the total X-ray luminosity is from larger radii, although the surface brightness decays as a function of radius.}
\label{fig:luminosity}
\end{figure}

The resultant X-ray luminosity of the gas halo is an important constraint for an AGN feedback model \citep[e.g.][]{2015MNRAS.449.4105C,2010MNRAS.406..822M}. \fref{fig:luminosity} shows the predicted X-ray cooling luminosity, integrated over all gas in the halo, from $0.5-7$\,keV. The luminosity  is calculated  using the same methods in \cite{2009A&A...508..751S,2018MNRAS.478.3544R}, in which the cooling curve is calculated  for the photospheric solar abundances \citep{2003ApJ...591.1220L}, using the spectral analysis code SPEX \citep{1996uxsa.conf..411K} and scaled according to the local  hydrogen, helium, and metal mass fractions. 
The shaded regions indicate the observed  X-ray luminosities in \cite{2002ApJ...567..716R} and \cite{2006ApJ...648..956S} for halos with $m_{\rm halo}\sim 0.7-1.5 \times 10^{14} M_\odot$.
Runs which quench by violently ejecting gas strongly suppress their X-ray luminosities, as does the ``uniform'' turbulent stirring run (owing to its suppression of gas densities everywhere in the halo). But interestingly, other runs with suppressed SF/cooling flows maintain X-ray luminosities just a factor $\sim 1.5-3$ lower, well within the observed range \citep{2002ApJ...567..716R,2006ApJ...648..956S,2006MNRAS.366..624B,2013ApJ...776..116K,2015MNRAS.449.3806A}. This is because a large portion of the total X-ray luminosity is from larger radii, although the surface brightness decays as a function of radius.

\subsubsection{Turbulent Mach number} \label{S:mach}

\begin{figure*}
\centering
\includegraphics[width=16cm]{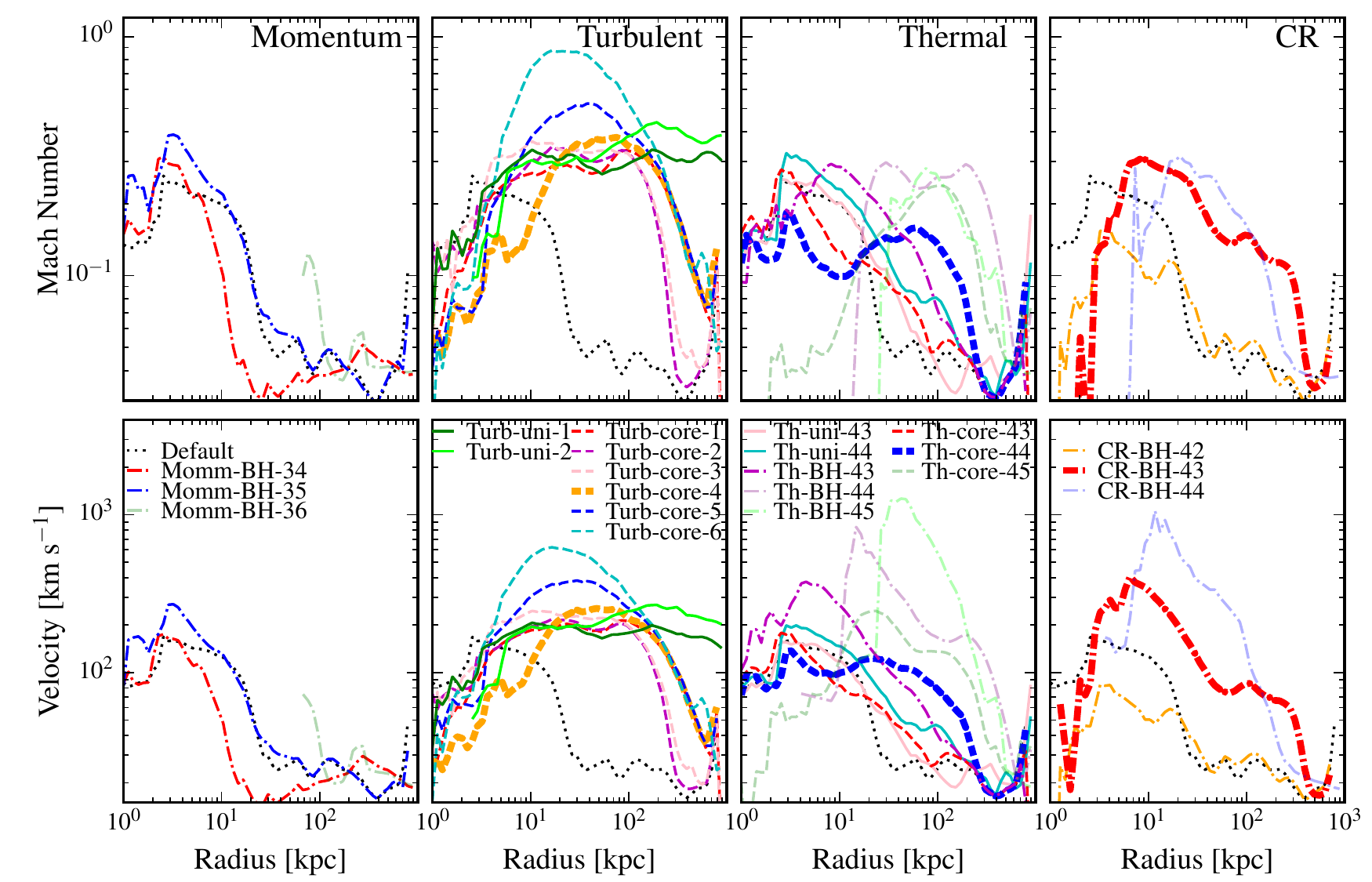}
\caption{{\em Top:} 1D rms Mach number ($v_{\rm turb}/\sqrt{3} v_{\rm thermal}$, in gas with $T>10^{7}\,$K, averaged over the last 100\,Myr of the runs) as a function of radius for the runs in \fref{fig:temp}. {\em Bottom:} 1D rms velocity dispersion $v_{\rm turb}/\sqrt{3}$. The `Default' run has turbulence driven by a combination of thermal instability and stellar feedback. 
Weaker momentum/thermal/CR input does not alter this much; when those inputs become ``explosive'' (see \fref{fig:temp}), strong shocks appear as jumps in $v_{\rm turb}$ up to $\gtrsim 1000\,{\rm km\,s^{-1}}$. Modest CR injection or distributed thermal injection contribute $\sim 200\,{\rm km\,s^{-1}}$ bulk motions at $r\gtrsim 10\,$kpc. Turbulent stirring runs (by construction) produce Mach $\sim 0.2-0.4$ turbulence over the radii of the chosen kernel, although the strongest runs (e.g.\ `Turb-core-6') exceed Mach $\gtrsim 1$. 
}
\label{fig:mach}
\end{figure*}

\fref{fig:mach} shows the rms 1D turbulent velocity, defined as $v_{\rm turb}/\sqrt{3}$, and the 1D Mach number for gas hotter than $10^7$K as a function of radius, averaged over the last 100 Myr of the runs. Radial momentum injection does not alter turbulence much, as it primarily drives coherent motion; likewise for thermal injection when it is weak or spread over large radii. In the ``explosive'' regime of momentum/thermal/CR input, all drive strong outflows at up to $\sim 1000\,{\rm km\,s^{-1}}$, though the higher shocked-gas temperatures mean this corresponds to Mach$\sim 0.4$. At intermediate CR injection rates, appreciable but modest bulk motions are driven at $\gtrsim 10\,$kpc. 

By construction, turbulent stirring boosts turbulent velocities where injected. The maximal turbulent velocities reach $\sim 200-400\,{\rm km\,s}^{-1}$ (Mach $\lesssim 0.5$) in the `Turb-uni’ and `Turb-core-1-5' runs, broadly consistent with observations \citep{2016Natur.535..117H,2018PASJ...70....9H}, but towards the higher end of the allowed range, while `Turb-core-6' exceeds Mach $>1$. 



\subsection{Cooling time and gas stability} \label{S:coolingtime}

\begin{figure*}
\centering
\includegraphics[width=16cm]{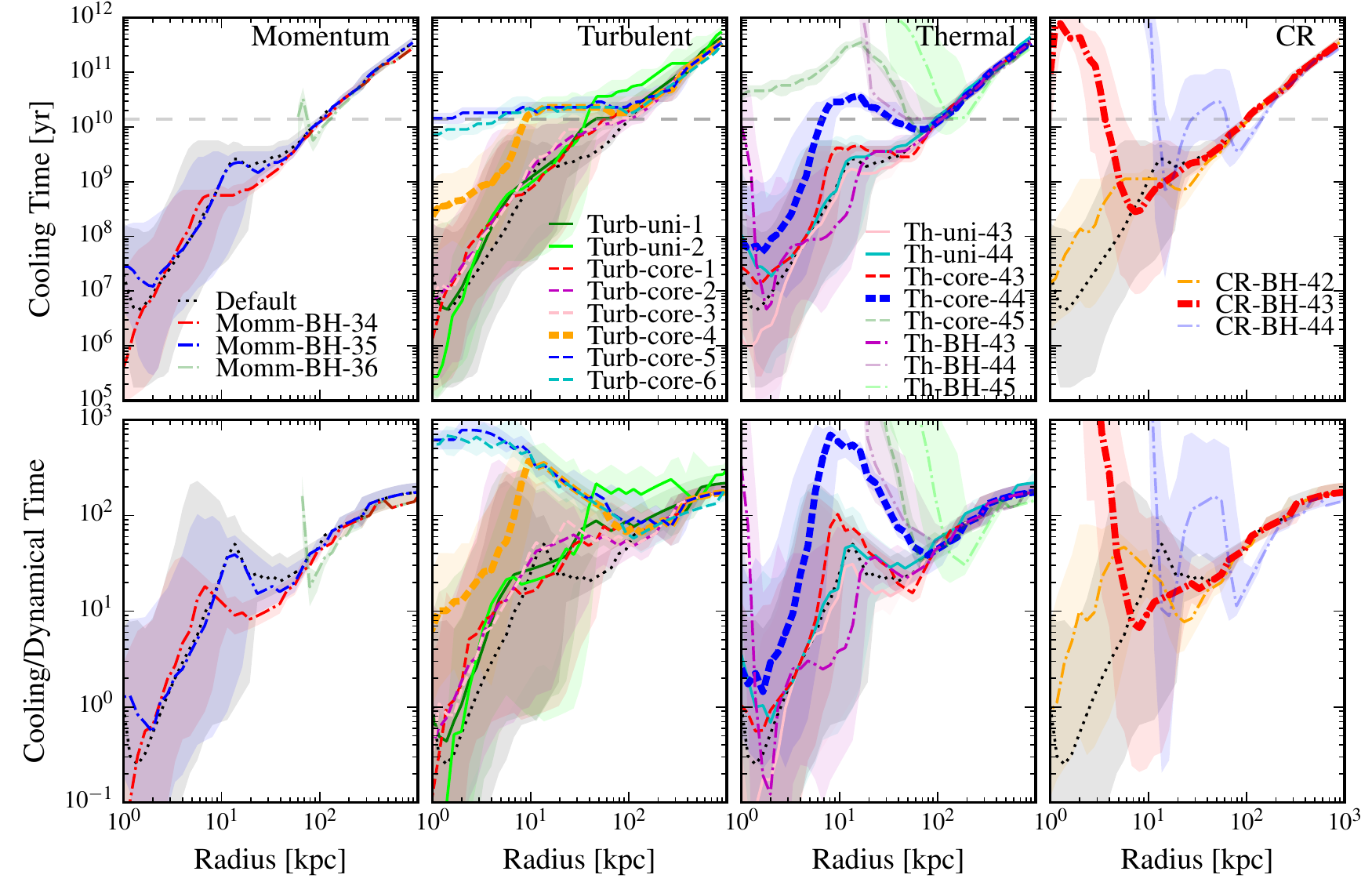}
\caption{{\em Top:} Gas cooling time ($\tau_{c}\equiv E_{\rm thermal}/\dot{E}_{\rm cooling}$) versus radius (averaged in the last $100\,$Myr of the runs in Fig.~\ref{fig:temp}). Gray dashed line labels the Hubble time. {\em Bottom:} Cooling time over dynamical time ($\tau_{d} \equiv (r^3/GM_{enc})^{1/2}$). With weak injection these are not strongly modified. In explosive cases the cooling time interior to the explosive shock is enormous (the ``cutoff'' to zero reflects cases where there is no gas in the relevant temperature range inside some radius). CRs and turbulent stirring suppresses cooling primarily by suppressing core gas densities; regions with boosted $\tau_{c}$ correspond to regions with strong stirring. If the stirring exceeds $\gtrsim2\times10^{41} {\rm erg\,s}^{-1}$ within 100 kpc   (`Turb-core-5 \& 6'), or CR injection exceeds $\gg 10^{43}\,{\rm erg\,s^{-1}}$, the gas has $\tau_{c}>t_{\rm Hubble}$ and $\tau_{c}/\tau_{d} \gtrsim 100$ -- this is an excellent predictor of when the system will resemble a non-cool-core-cluster. 
}
\label{fig:coolingtime}
\end{figure*}

\fref{fig:coolingtime} shows the cooling time ($\tau_{c}\equiv E_{\rm thermal}/\dot{E}_{\rm cooling}$) versus radius for gas hotter than $10^{5}$\,K. Momentum injection does not affect the cooling time strongly. Even in our highest momentum flux run ($\sim10^{36} {\rm g\,cm\,s}^{-2}$), where everything within 70 kpc is blown away, the cooling time at even larger radii still remains very similar to the `Default' run.

On the other hand, turbulent stirring, which effectively  suppresses the gas density, can also  suppress the cooling rate (through turbulent mixing and pressure support). The regions with boosted cooling time roughly coincide with the regions with strong stirring. When the  stirring injects  $\sim \times10^{41} {\rm erg\,s}^{-1}$ (`Turb-core-4') within 100 kpc, the average cooling time  of gas with $T>10^5$\,K beyond 10 kpc is  boosted to $\gtrsim 10\,$Gyr. In `Turb-core-5 and 6', almost all the gas becomes stably non-cooling, consistent with their resulting non-cool-core halo properties.

Thermal heating can  significantly boost the cooling time as long as the kernel of injection is small enough ( only the `Th-BH' runs and `Th-core' runs). The  increase of the cooling time basically follows the increase in temperature discussed in \sref{S:temp}.

CR injection  boosts the cooling time  {significantly within}  $r\sim10$ kpc, owing to lower densities and higher temperatures inside these radii, when the injection rate is $\gtrsim 10^{43} {\rm erg\,s}^{-1}$. 

The ratio of cooling time to dynamical time ($\tau_{c}/\tau_{d}$, with $\tau_{d}\equiv(r^3/GM_{enc})^{1/2}$) is also plotted as an indication of gas stability.  The runs suffering from the most severe cooling flows in our suite (`Default', `Momm-BH-34,35', `Th-uni-43,44', `Th-core-43', `Th-BH-43' and `CR-BH-42' ) have an extended region within 100 kpc at $\tau_{c}/\tau_{d}\lesssim 20$. In the runs with SFRs suppressed to $\lesssim 1 {\rm M}_\odot {\rm yr}^{-1}$ (`Turb-core-4', `Th-core-44' and `CR-BH-43'), most of the gas within this radius has $\tau_{c}/\tau_{d}>10$. In the runs which end up resembling non-cool-core clusters (`Turb-core-4 \& 5'), $\tau_{c}/\tau_{d}\gtrsim 100$ uniformly. 
Consistent with previous studies  \citep[e.g.][]{2012MNRAS.420.3174S,2015A&A...579A..62G,2017ApJ...845...80V}, we find that our simulations that avoid the cooling catastrophe and also produce ``realistic'' cool-core profiles have $\tau_c/\tau_d \gtrsim 10$.

\subsection{Energy input vs. Cooling} 
\label{S:energyvscooling}

\begin{figure*}
\centering
\includegraphics[width=17cm]{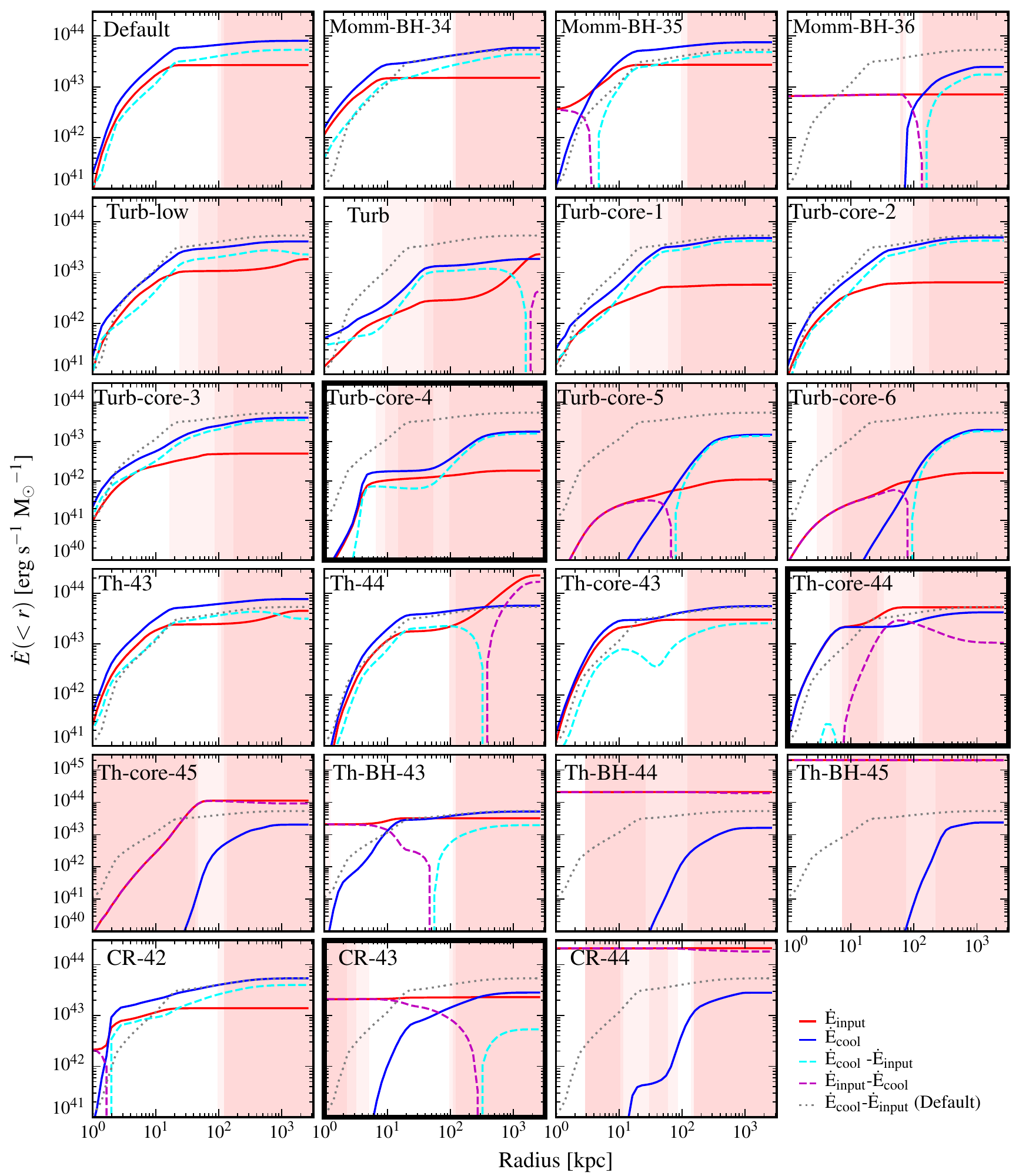}
\caption{Cumulative (integrated inside $<r$) cooling rate ($\dot{E}_{\rm cool}$), total feedback energy input rate ($\dot{E}_{\rm input}$), and difference (net loss/gain), in the runs from \fref{fig:temp} averaged over their last $100\,$Myr. Progressively darker red shading indicates regions where $20/50/80\%$ of gas (above $T>10^{5}$\,K) has cooling times greater than the Hubble time.
Weak input produces little change; ``explosive'' runs quench by dramatically lowering heating and core gas densities. Several of the turbulent runs suppress cooling significantly, ensuring $\tau_{c}>t_{\rm Hubble}$, without ``exploding.'' This is also seen in intermediate CR runs (`CR-BH-43'), more or less independent of the injection kernel. In thermal runs this requires an injection kernel and energy fairly carefully matched to the cooling radius/energy.
}
\label{fig:in_out}
\end{figure*}

In \fref{fig:in_out}, we compare cooling rates, energy input rates, and net energy gain/loss of each run, integrated within a radius $r$. Here ``energy input'' sums stellar feedback (adding SNe and stellar mass-loss kinetic luminosities) plus the input from our analytic injection models. We also show where gas (above $10^{5}$\,K) has cooling times exceeding the Hubble time.

 Direct thermal heating, as expected, suppresses cooling in the core region only if the injected heating rate is larger than cooling: this is why uniform or large-kernel heating is inefficient (energy is ``wasted'' at large radii). When highly-concentrated, this tends to result in explosive behavior, which reduces the cooling rate further out not by direct heating but by ejecting the halo baryons. The only thermal heating run with heating roughly matched to cooling over the extended cooling region is the (intentionally fine-tuned) `Th-core-44' run. 
 
 Akin to the thermal runs, `CR-BH-42' does little, `CR-BH-44' is explosive, while `CR-BH-43' is able to maintain quasi-stable equilibrium. A key difference is (as we show below) this comes primarily from pressure support, where the CR pressure profile (if diffusion is fast and the injection rate is constant, and losses are negligible) is essentially a steady-state $p_{\rm cr} \sim \dot{E}_{\rm cr} / 12\pi\,\kappa\,r$. This makes the predictions less sensitive to small variations in the cooling rates or gas densities. 
 
 Turbulent stirring can suppress cooling rates significantly without becoming ``explosive'' and with significantly lower energetic ``cost.'' We discuss the mechanisms for this in \sref{dis:turb}.



\subsection{The Rejuvenation of Non-cool-core Clusters, and Role of Feedback from Old Stellar Populations}

Given that `Turb-core-5' and `Turb-core-6' evolve from cool-core to non-cool-core in a relatively ``gentle'' manner, a natural question to ask is whether the halo will become cool-core again if the turbulent stirring is turned off. It turns out that rejuvenation does not necessarily occur, at least in these idealized simulations (remember, our simulations are non-cosmological, so do not include {\em new} gas accreting into the halo). 

\begin{figure}
\centering
\includegraphics[width=8.5cm]{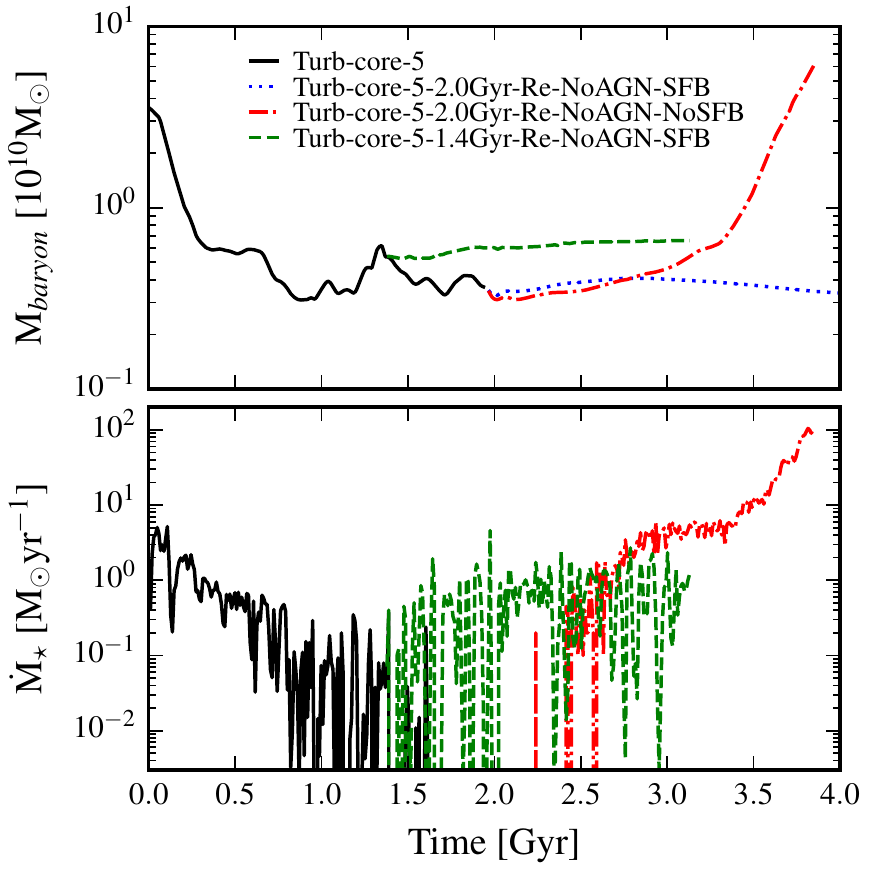}
\caption{Testing ``rejuvenation.'' We re-start run `Turb-core-5' which transitions from cool-core to non-cool-core cluster, at either 1.4 Gyr (`Re-1.4Gyr-SFB') or 2.0 Gyr (`Re-2.0Gyr-SFB'), keeping all physics identical but turning off the turbulent ``stirring'' at the time of restart. We compare the baryonic mass within $r<30\,$kpc (excluding pre-existing stars; {\em top}) and SFR ({\em bottom}) as Fig.~\ref{fig:sfr}. The earlier restart ``rejuvenates'' after additional energy injection is disabled, and $\dot{M}_{ast}$ slowly re-grows over $\sim$\,Gyr timescales. The later restart fails to rejuvenate, as in the intervening time, continued driving has lowered the core gas density to the point where stellar feedback from old stars (Ia \&\ AGB) can keep it hot. We confirm the latter by re-running the 2.0 Gyr restart without this feedback (`Turb-core-5-2.0Gyr-Re-NoAGN-NoSFB'), which now rejuvenates. 
}
\label{fig:sfr_rj}
\end{figure}

\begin{figure}
\centering
\includegraphics[width=8.5cm]{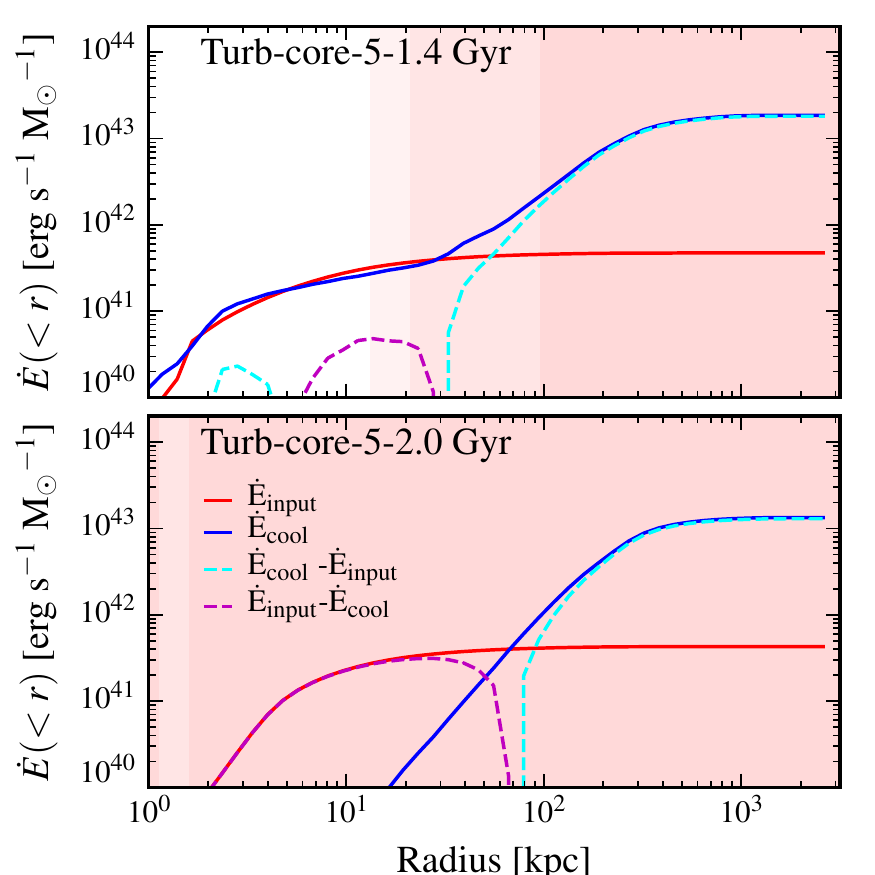}
\caption{Cumulative cooling rate versus energy input (as Fig.~\ref{fig:in_out}, but including {\em only} stellar feedback from old stellar populations in the $\dot{E}_{\rm input}$ budget), of the ``re-started'' runs in \fref{fig:sfr_rj}. In the earlier restart, the higher core densities allow cooling and rejuvenation after the turbulent injection is de-activated. In the later restart, the stellar injection is identical (it comes from the same old stars) but the gas density has been further lowered by the injection to the point where old stars can maintain quenching for a Hubble time.}
\label{fig:in_out_rj}
\end{figure}

We test this by restarting the `Turb-core-5' run from the 1.4\,Gyr point and the 2.0\,Gyr point, removing our injection (keeping e.g.\ stellar feedback and all other physics, however). As shown in \fref{fig:sfr_rj}, the `1.4 Gyr' run rejuvenates (core baryonic mass slowly grows and star formation reoccurs) while the `2.0 Gyr' one does not. The reason is that once the density is lowered to a (very low) point where the residual steady-state energy input from type 1a SNe and AGB winds surpasses cooling, the halo remains quenched for a Hubble time. \fref{fig:in_out_rj} shows the same comparison of \fref{fig:in_out} for `Turb-core-5' at 1.4 Gyr and 2.0 Gyr, but includes only the stellar feedback contribution in ``energy input.'' It is clear that at 1.4 Gyr there is still an extended region ($r\lesssim 30$ kpc) with sufficiently dense gas that stellar feedback from old stars alone (SNe Ia \&\ AGB mass-loss) can only marginally balance cooling, while by 2.0 Gyr, the density has been depleted to the point where the old-star stellar feedback (which is basically identical) now totally surpasses cooling. 

If we restart this 2.0 Gyr run without stellar feedback from old stars (disabling Ia's and AGB mass-loss), then it does rapidly resume SF and ``rejuventate.'' We have confirmed it is the Ia population which dominates the energy injection and results here. But in either case, it appears that stellar feedback can aid in {\em maintaining} quenched systems, but only once they are well into the non-cool-core stage with especially-depleted central gas densities.

We focused on this case because it was only marginally a non-cool-core cluster. In every simulations that produces ``explosive'' quenching, the central gas densities are extremely low (much lower than our 2.0 Gyr run here) and so, unsurprisingly, rejuvenation never occurs.

\section{Results As a Function of Halo Mass} \label{S:halo mass}

We now explore models in lower-mass halos {\bf m12} and {\bf m13} at $M_{\rm halo}\sim 10^{12}$ and $10^{13}\,{\rm M}_{\odot}$, respectively (see \tref{tab:ic}). We focus our attention on models motivated by those that at least seem plausibly ``successful'' (able to have some effect, but also not obviously in gross violation of observational constraints) -- this includes variations of the turbulent stirring ``core-kernel'' runs, cosmic ray injection with appropriate energetics, and thermal heating with an appropriate-scaled spatial kernel and energy scale. The survey at high resolution is listed in \tref{tab:run2}, though we have run additional low-resolution tests of broader parameter space to confirm our intuition from the {\bf m14} survey continues to hold. 

For the thermal and CR injection cases, we scale the input energy from the m14 `Th-core-44' and `CR-BH-43' runs according to the total cooling rate of the halo. For turbulent stirring, we scale the characteristic wavelength of the stirring in Fourier space ($\lambda$) and kernel size ($r_k$) from the m14 `Turb-core-4' run according to the viral radius, and the amplitude of particle acceleration according to the circular velocity at the kernel size: $a r_k \sim v_c^2 \sim GM_{\rm enc}/r$.
For the {\bf m12} case, the above scaling makes the kernel very narrow and confines the stirring or energy injection to the disk, so we also included {\bf m12} runs with a wider kernel.

\setlength{\tabcolsep}{4pt}
\begin{table}
\begin{center}
 \caption{Physics variations (run at highest resolution) in our survey of lower-mass ({\bf m12} \&\ {\bf m13}) halos}
 \label{tab:run2}
\resizebox{8.5cm}{!}{%
 \begin{tabular}{cc|cc|c}
 \hline
\hline
Model                       &$\lambda$ &$\dot{E}_{tot}$  &$\dot{P}_{tot}$  &kernel\\
                            &(kpc)      &(erg s$^{-1}$)   &(g cm s$^{-2}$)  & (r in kpc)\\
\hline 
\mc{\mr{m12-Turb-core}}     &\mr{25}   &\mr{5.5-8.4 e39} &\mr{4.0-4.2 e32} &\mc{$a_{r<20}\sim3\exp(-(r/ 15.8)^2)$}\\
&&&                                                                         &\mc{$a_{r>20}\sim \exp(- r/39.6)$   }\\   
\mc{\mr{m12-Turb-core-wide}}&\mr{25}   &\mr{4.7-9.3 e39} &\mr{4.6-4.9 e32} &\mc{$a_{r<50}\sim3\exp(-(r/ 39.6)^2)$}\\
&&&                                                                         &\mc{$a_{r>50}\sim \exp(- r/100)$   }\\   
\mc{\mr{m13-Turb-core}}     &\mr{57}   &\mr{13-7.5 e39}  &\mr{10-3.8 e32}  &\mc{$a_{r<50}\sim3\exp(-(r/ 40)^2)$}\\
&&&                                                                         &\mc{$a_{r>50}\sim \exp(- r/100)$   }\\   
\mc{\mr{m14-Turb-core-4}}   &\mr{120}   &\mr{5.7-5.9 e41} &\mr{3.1-2.0 e34} &\mc{$a_{r<100}\sim3\exp(-(r/ 79)^2)$}\\
&&&                                                                         &\mc{$a_{r>100}\sim \exp(- r/200)$   }\\   
\hline
\mc{m12-Th-core-43}         &-          & 1.3-1.4 e43     & -               &$\dot{E}\propto\exp(-(r/6)^2)$\\         %
\mc{m12-Th-core-43-wide}    &-          & 1.3-2.5 e43     & -               &$\dot{E}\propto\exp(-(r/14)^2)$\\         %
\mc{m13-Th-core-43}         &-          & 17 -7.8 e42     & -               &$\dot{E}\propto\exp(-(r/14)^2)$\\         %
\mc{m14-Th-core-44}         &-          & 2.0-0.5 e44     & -               &$\dot{E}\propto\exp(-(r/30)^2)$\\         
\hline
\mc{m12-CR-BH-42}           &-          & 1.3e42          & -               &BH neighbour\\                               
\mc{m13-CR-BH-42}           &-          & 1.8e42          & -               &BH neighbour\\                               
\mc{m14-CR-BH-43}           &-          & 2.1e43          & -               &BH neighbour\\                               
\hline 
\hline
\end{tabular}
}
\end{center}
\begin{flushleft}
Partial list (including just simulations at ``production'' resolution) of runs in halos {\bf m12} and {\bf m13}. Style is identical to Table~\ref{tab:run}, but we add one column $\lambda$, denoting the wavelength of the turbulent driving modes. We focus only on models which were successful without being ``explosive'' in the {\bf m14} suite, and scale the energetics and kernel sizes with the cooling luminosities and virial radii, respectively. We have run additional low-resolution tests akin to the suite in Table~\ref{tab:run} to confirm much larger/smaller injection produces similar results to what is seen there. 
\end{flushleft}
\end{table}
\setlength{\tabcolsep}{6pt}

\begin{figure}
\centering
\includegraphics[width=8.5cm]{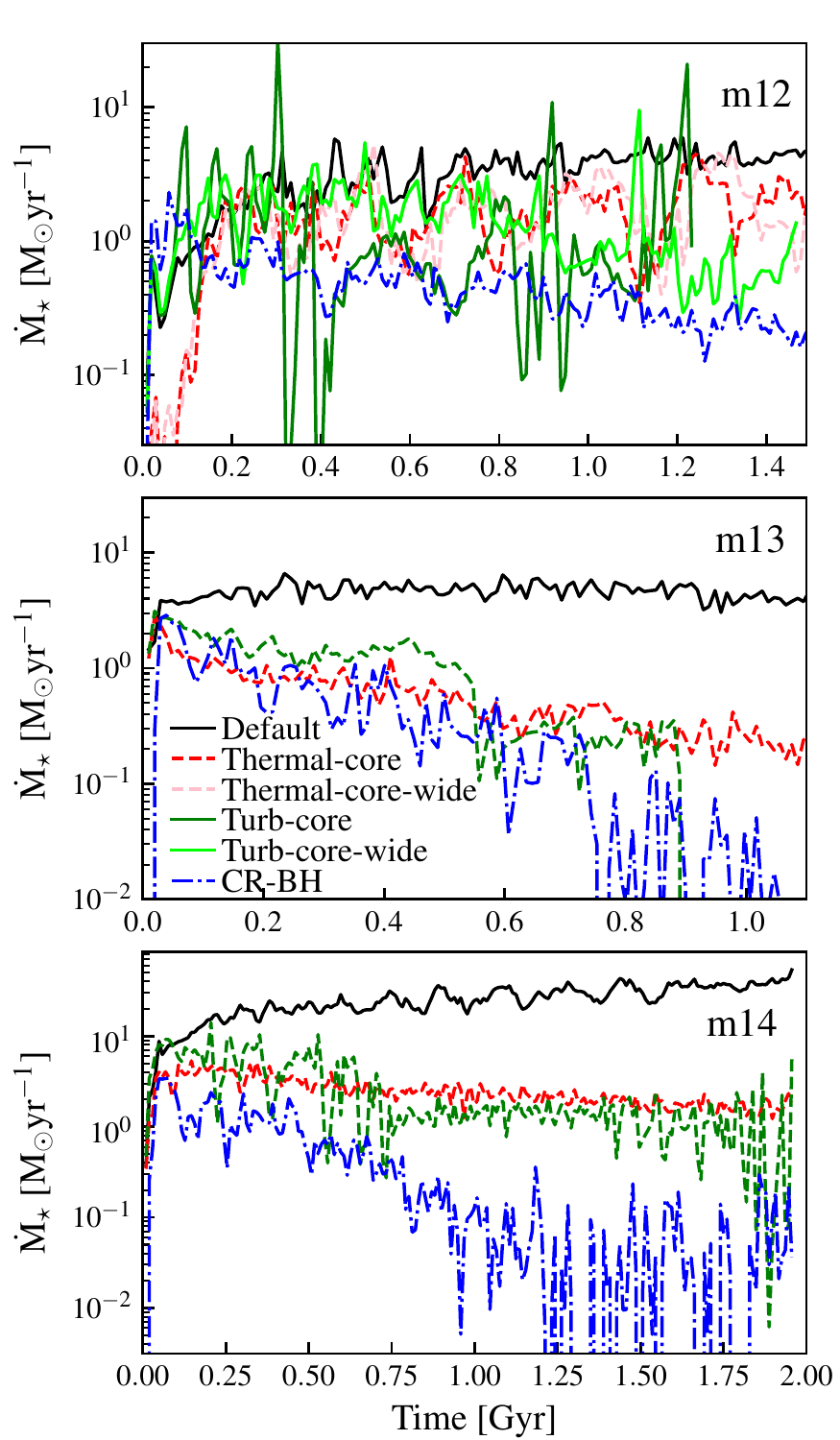}
\caption{Galaxy SFRs (as Fig.~\ref{fig:sfr}) in our suite of simulations of different mass halos (\tref{tab:run2}). The ``wide'' runs are only for m12.  In {\bf m13} and {\bf m14}, turbulent stirring within the halo scale radius, or CR injection with appropriate energies, can quench, as can somewhat fine-tuned thermal energy injection. 
{\bf m12} is more unstable and we find no thermal-heating solutions that quench without explosive ejection of halo baryons. Also in {\bf m12} the `Turb-core' run confines stirring to $\sim 16$\,kpc, effectively ``churning'' the galactic disk and producing the bursty star formation; this disappears with a more extended stirring (`Turb-core-wide').
}
\label{fig:sfr_m1213}
\end{figure}

\begin{figure*}
\centering
\includegraphics[width=18cm]{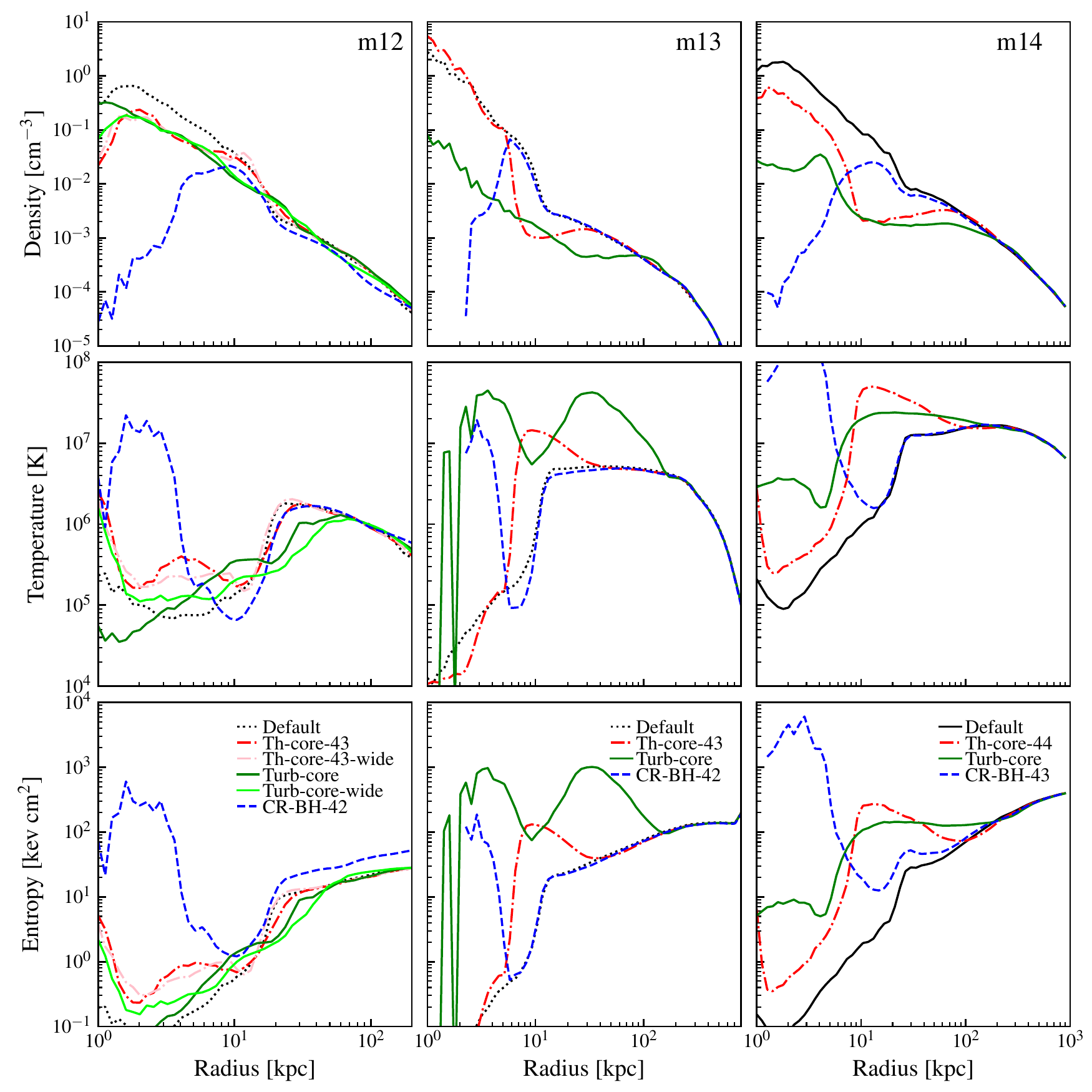}
\caption{Density, temperature, \&\ entropy profiles (as Fig.~\ref{fig:temp}) averaged over the last $\sim100\,$Myr of each run for the runs in \fref{fig:sfr_m1213}. Even fine-tuned thermal heating produces negative temperature gradients in {\bf m13} \&\ {\bf m14} and has little effect in {\bf m12}. Turbulent stirring significantly depresses the density of {\bf m13} (and raises its temperature), resembling the stronger-stirred {\bf m14} cases, but a weaker stirring amplitude alleviates this. CR injection suppresses the core density inside the central few kpc (leading to mostly hot gas inside this radius), but leaves a positive temperature profile and intact density profile outside $r>5-10$\,kpc. 
} 
\label{fig:etd_range}
\end{figure*}

\begin{figure*}
\centering
\includegraphics[width=18cm]{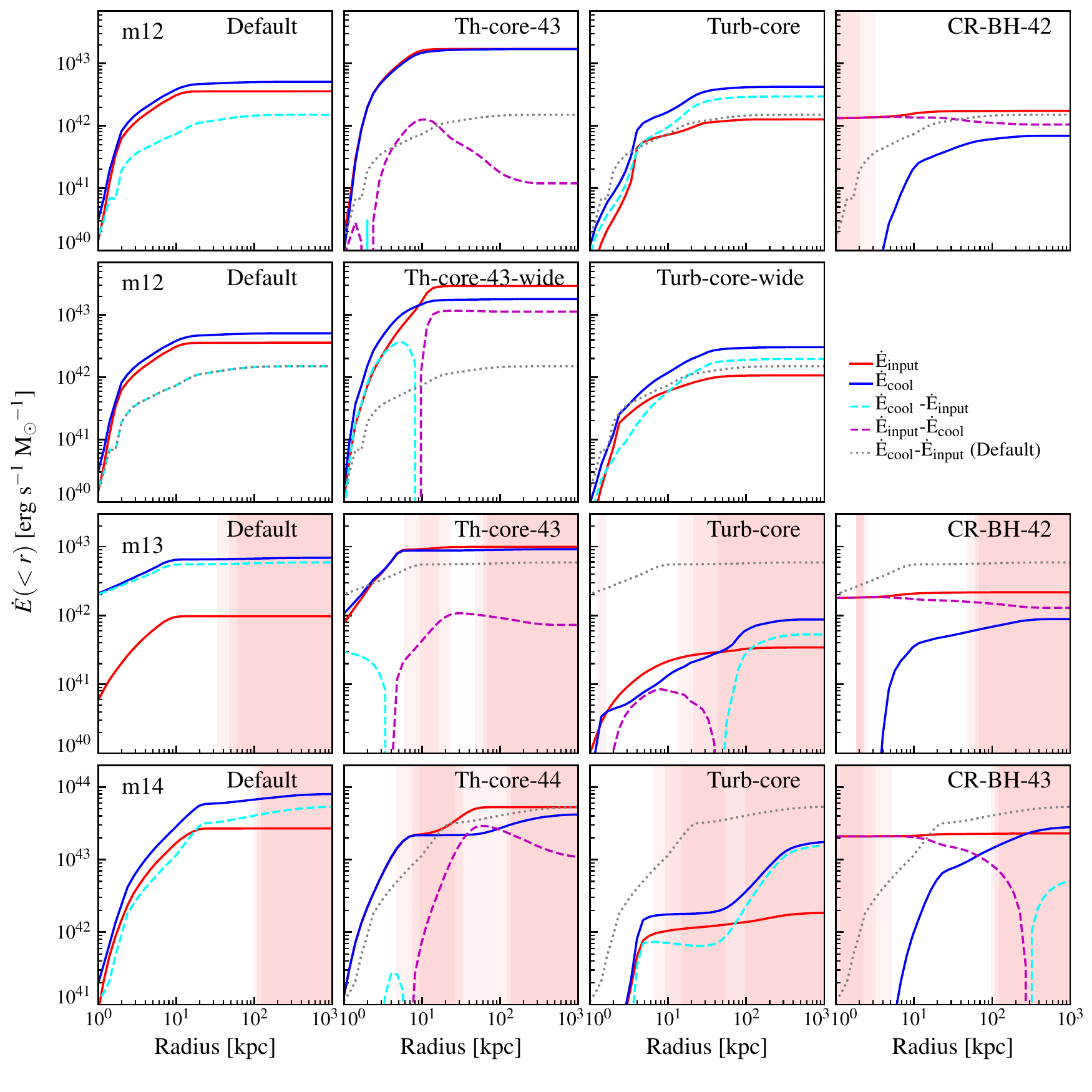}
\caption{Cumulative energy input vs.\ cooling (as Fig.~\ref{fig:in_out}) for the simulations in Fig.~\ref{fig:sfr_m1213}. Turbulent stirring significantly decreases cooling in {\bf m13} (owing to lower densities in Fig.~\ref{fig:etd_range}; this is sensitive to the injection rate), but only suppresses cooling in the core in {\bf m12}. Thermal heating can marginally balance cooling in {\bf m13} and {\bf m14} which have hotter, more stable gaseous halos, but in {\bf m12} heating puts more dense+enriched gas at $\sim 10^{4}-10^{5.5}$\,K, increasing its cooling rate. CR injection substantially suppresses cooling rates in all cases.}
\label{fig:inout_range}
\end{figure*}

The resulting SFRs are plotted in \fref{fig:sfr_m1213}. Turbulent stirring and CR injection quench all halos, while thermal heating is less efficient in {\bf m12}.
\fref{fig:etd_range} shows the density, temperature, and entropy profiles, while \fref{fig:inout_range} compares the energy injection to cooling luminosities as a function of radius. 

Thermal heating has similar effects in {\bf m13} and {\bf m14}: the galaxies are quenched but inevitably have a mild negative temperature gradient. The energy input of both matches the cooling in an extended region (by construction). However, in the {\bf m12} case, the cooling rate is actually {\em boosted} by the additional thermal heating (because the virial temperature is low and amount of gas at $\sim 10^{4}-10^{5}$\,K is large, this pushes gas higher on the cooling curve), so the effect is much weaker. In lower-resolution tests, this is not remedied by increasing the thermal energy injection rate: because of the more violent thermal instability, all our thermal-heating runs in {\bf m12} either produce no effect, or violent explosion of the entire halo.


Naively scaling the turbulent ``stirring scale'' with the viral radius (and strength with the circular velocity) quenches {\bf m13}, but the stirring at large radii $>50\,$kpc is too strong and a substantial gas mass is thrown out of the halo completely, lowering the density and X-ray luminosity, while the stirring at small radius makes the halo non-cool-core (akin to stronger stirring cases in {\bf m14}). This can be moderated, as expected, with a somewhat weaker stirring. In {\bf m12} these naive scalings lead to stirring confined to $\sim 16$\,kpc, which effectively stirs the galactic disk and ballistically ``launches'' the whole disk into fountains, which produce a violently bursty star formation history. Obviously this is not realistic: increasing the ``stirring kernel'' size to $\sim 40$\,kpc (`m12-turb-core-wide') produces a much smoother low SFR and stable cool-core structure with slightly-lower core densities and cooling rates.


CR injection successfully and ``smoothly'' quenches {\bf m12} and {\bf m13}. In {\bf m13} the overall cooling rate is eventually also suppressed significantly as the gas within the central few kpc is ejected.


\section{Discussion: How Do Different Physics Quench (or Not)?} 
\label{s:discussion}

The only injection models that result in a semi-stable quenched galaxy are thermal heating with a Gaussian kernel chosen in the correct energy and size range, turbulent stirring confined to radii below the halo scale radius, and cosmic ray heating in the correct energy range. We briefly discuss how each surveyed model in \sref{S:physics} operates.

\subsection{Radial Momentum injection}

The actual kinetic energy (even if it all thermalized) of the momentum-injection runs is less than the cooling luminosity.\footnote{The total kinetic energy input roughly matches the total cooling rate  at the begin of the most explosive ``Momm-BH-36'' case, but soon become sub-dominant.} At low injection rates this just stirs small-scale turbulence/fountains (e.g.\ in ``Momm-BH-35,'' 40\% of the momentum and 20\% of the energy  input is used to decelerate in-falling gas, and $\sim 1/2$ of the gas acted directly upon by the nuclear stirring still forms stars). At higher injection rates it acts by dynamically altering halo structure, ejecting material from the core. Without truly enormous energy input this is eventually decelerated in the outer halo, but in e.g.\ ``Momm-BH-36'' almost all the gas within $\lesssim 70\,$kpc is ejected. 

Previous studies have similarly noted that pure isotropic kinetic input tends to fall into burst-quench cycles where either it fails to alter the cooling flow, or explosively ejects all the gas in the cooling radius \citep{2009ApJ...699...89C,2010ApJ...711..268S}. One alternative is to inject energy in a completely different form (discussed below). A second is to inject momentum at larger radii (distributing it away from the center), in a spatially-localized-but-time-dependent manner (i.e.\ not in a simple radially-outward-moving shell, which simply repeats this problem on larger scales; see \citealt{2011MNRAS.411..349G,2014ApJ...789...54L,2016ApJ...818..181Y,2017MNRAS.472.4707B} for kinetic jets) -- this is much closer in practice to our ``turbulent stirring'' runs below. A third alternative is to invoke a mix of isotropic kinetic feedback and thermal feedback \citep[as in][]{2010ApJ...717..708C,2013MNRAS.428.2885D,2017MNRAS.465.3291W,2018MNRAS.473.4077P}, a possibility we discuss below.

\subsection{Thermal heating}

In pure thermal heating models, nothing changes unless the overall heating rate is larger than the cooling rate. However, unless these are carefully balanced, this tends to produce a negative temperature gradient in direct contradiction with observed systems \citep{2002ApJ...567..130B,2006ApJ...638..659M}, and can drive explosive behavior which removes most of the gas in the halo. As a result, we must tune the energy input to match the cooling rate. We must also tune the injection radius to match the cooling radius, or else the energy is either ``wasted'' on gas at large radii (not cooling efficiently), or it excessively heats gas in the center driving Sedov-Taylor blastwaves that heat gas to very high temperature,
eject gas in the central halo, produce negative temperature gradients, and strong shocks. 

This is also consistent with previous studies that have repeatedly found nuclear energy injection alone tends to either fail to quench, or violently eject far too much gas from halos \citep{2014MNRAS.445..175G}. 
The alternatives are typically to invoke either (1) fine-tuning, or (2) some mix of other feedback mechanisms.

\subsection{Turbulent stirring}
\label{dis:turb}
In almost all of our turbulent stirring runs which produce suppressed cooling flows, the turbulent energy injection rate is much lower than the total cooling rate (especially pre-turbulence) -- in other words, thermalized turbulent energy ``heating'' gas is not the dominant channel. We have also directly confirmed this by measuring the turbulent damping rate and comparing it to cooling. More important, turbulence mixes gas to larger radius, which (a) lowers the central density, (b) lowers the density of ``up-welling'' parcels (lowering their cooling rates), and (c) mixes them with hot gas (providing a form of ``bulk conduction''). Together this lowers the effective cooling rate by an order of magnitude in the  region with cooling times shorter than the Hubble time.

\begin{figure}
\centering
\includegraphics[width=8.5cm]{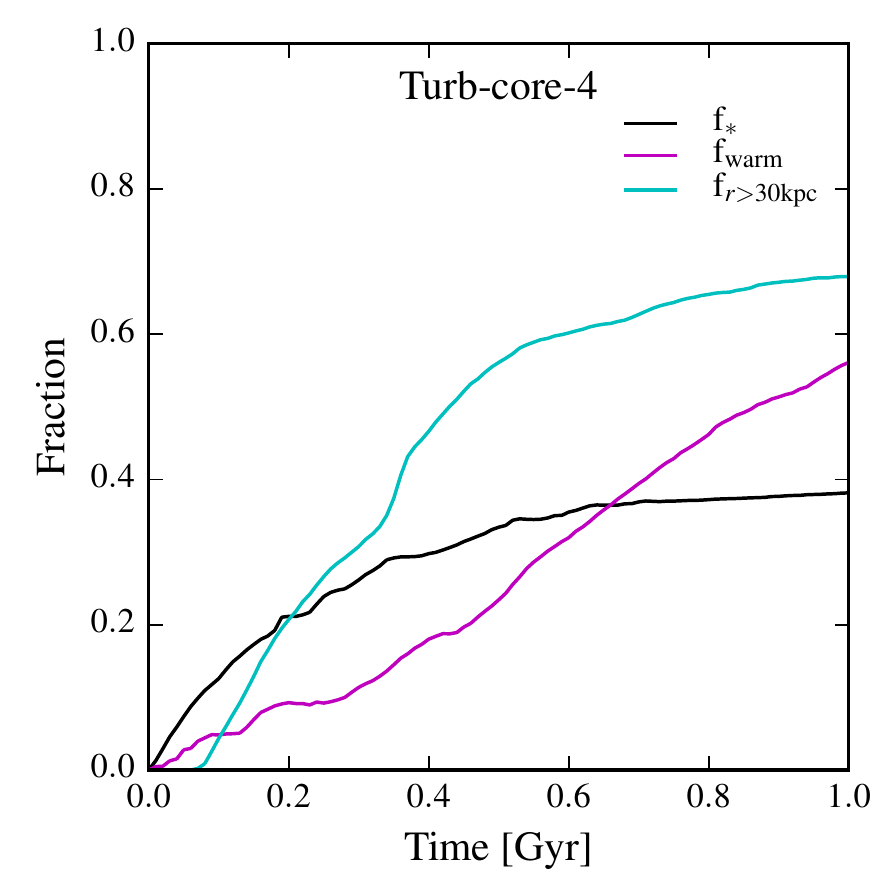}
\caption{An example of what happens to rapidly-cooling gas in the turbulent stirring runs which suppress cooling flows. Here in `Turb-core-4' from \fref{fig:temp}, we track gas which is cold ($T<8000$K) and dense ($n>1 {\rm cm}^{-3}$) at an early time ($20$\,Myr) and follow its evolution. We first follow how much of the gas  forms stars -- this stabilizes as $\sim 40\%$ at late times. The majority of the gas is mixed to larger radii (here $>30\,$kpc), and becomes warm ($T>10^{5}\,$K). 
}
\label{fig:mixing}
\end{figure}

Taking `Turb-core-4' for instance, \fref{fig:mixing} tracks the evolution of gas which is cold and dense at one initial time. After $\sim 0.5\,$Gyr, less than half remains cold and dense or forms stars: most is shifted to larger radius and mixed into with hot gas \citep[e.g.][]{2003ApJ...596L.139K,2006ApJ...645...83V,2010ApJ...712L.194P,2010ApJ...713.1332R,2011MNRAS.414.1493R,2014MNRAS.443..687B}. 

\subsection{Cosmic Ray Injection}

\begin{figure*}
\centering
\includegraphics[width=16cm]{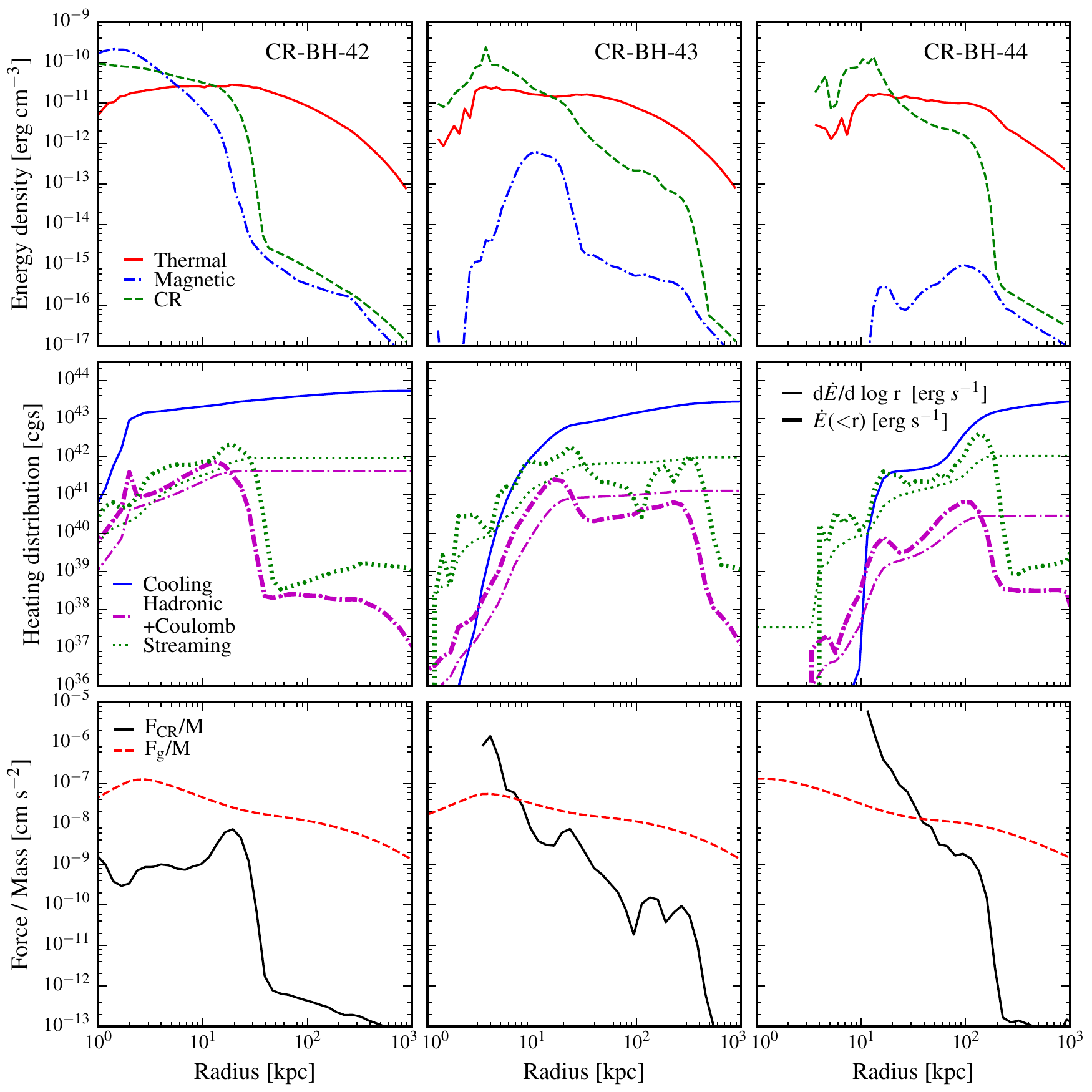}
\caption{Comparison of energetics in our {\bf m14} cosmic ray runs. 
{\em Top:} CR, magnetic, and thermal energy densities (averaged in spherical shells in the last 100 Myr of each run), for the three {\bf m14} CR runs (\tref{tab:run}). CR energy is non-negligible within $<30\,$kpc in each. 
{\em Middle:} Comparison of differential per-unit-radius ($d\dot{E}/d\log{r}$) and cumulative ($\dot{E}(<r)$) gas cooling rates versus CR ``heating'' rates. The latter includes collisional (hadronic+Coulomb) and streaming losses which transfer energy from CRs to thermal gas energy. CR heating is always much smaller than gas cooling except where the gas is almost completely evacuated in the central few kpc (and $\dot{E}_{\rm cool}$ is extremely small). 
In CR-BH-42, which retains dense central gas where CR losses are large, $\sim 60\%$ of the injected CR energy is thermalized, but in the CR-BH-43/44 runs where the central few kpc are lower-density, only $\sim1-5\%$ of the CR energy is ever thermalized.
{\em Bottom:} Gravitational acceleration $F_{g}/M \approx \partial \Phi/\partial r$ vs.\ acceleration from the CR pressure gradient ($\rho^{-1}\,\partial P_{\rm cr}/\partial r$). CR pressure dominates and pushes material out from the central cooling-core, to larger-$r$ at larger $\dot{E}_{\rm cr}$. 
\label{fig:e_compare}}
\end{figure*}

Like turbulent stirring, CR injection provides another source of  non-thermal pressure support, so the gas density and the cooling rate can be suppressed without directly heating up the gas. 
As shown in \fref{fig:e_compare}, CR energy does contribute as an important pressure source, reaching at least equipartition to the thermal energy.
Moreover, CR diffusion spreads out the energy input and forms a quasi-steady-state isotropic pressure gradient even if all the CR energy is injected in the vicinity of the black hole. 


Fig.~\ref{fig:e_compare} shows that only a small fraction of the CR energy is thermalized  as the CRs propagate in the `CR-BH-43/44' models,\footnote{The total thermalized CR energy as indicated by the green and magenta lines in the second row of \fref{fig:e_compare} is less than $10\%$ of the total CR input.} and in all cases the heating from CRs is well below total cooling rates. This is expected: the timescale for CRs to lose energy to hadronic+Coulomb processes is $\sim 30\,{\rm Myr}\,(n/{\rm cm^{-3}})^{-1}$ while the diffusion timescale is $\sim r^{2}/\kappa_{\rm cr}$, so for our parameters losses in the core are only significant if its mean density exceeds $n \gtrsim 0.1\,{\rm cm^{-3}}\,(r_{\rm core}/10\,{\rm kpc})^{-2}$.\footnote{This explains why the run `CR-BH-42', which does not quench and maintains dense gas in the center, does lose a non-negligible fraction $\sim 1/2$ of its CR energy to collisional+streaming losses. For `CR-BH-43' and `CR-BH-44', the collisional loss is more significant initially, but it drops to a lower value after the core density is suppressed. The competition between CR energy and gas densities being larger at small $r$, and diffusion times longer at large $r$, also explains why the collisional+streaming losses have the broad radial structure seen in Fig.~\ref{fig:e_compare}.} However, in `CR-BH-43/44,' the temperature in the very center ($\lesssim 5-10\,$kpc) does become large: this owes to CR pressure gradients suppressing the nuclear gas density sufficiently so that the low-density gas is heated efficiently by stellar feedback from the bulge and CR streaming heating.

On the other hand, the CR pressure gradient in Fig.~\ref{fig:e_compare} is able to offset gravity. If losses are negligible and diffusion dominates transport, around a point source with constant $\dot{E}_{\rm cr}$, the equilibrium pressure profile (assuming CRs are a $\gamma=4/3$ ultra-relativistic fluid) is $P_{\rm cr} = \dot{E}_{\rm cr}/12\pi\,\kappa\,r$, which agrees well with the inner parts of our CR runs (outside the ``holes'' in  `CR-BH-43/44' within the central few kpc, where stellar feedback dominates).\footnote{At large radii, if the streaming is at a quasi-constant Alfv\'en speed, streaming will dominate over diffusion at $r \gtrsim \kappa_{\rm cr} / v_{\rm stream} \sim 30\,{\rm kpc}\,(\kappa_{\rm cr}/10^{29}\,{\rm cm^{2}\,s^{-1}})\,(30\,{\rm km\,s^{-1}} / v_{\rm stream})$, which also defines the radius where streaming losses $\propto v_{\rm stream}\,\partial P_{\rm cr}/\partial r$ will be largest. Note in the simulations here the ``cutoffs'' in the CR profiles at $r \sim 100-1000\,$kpc owe to the simulations only having finite time for CRs to propagate from the nucleus to large radii.} Comparing this to the gravitational force we have:
\begin{align}
\frac{F_{\rm CR}}{F_{\rm G}}=&\frac{1}{3\rho}\frac{\partial e_{\rm CR}/\partial r}{GM_{\rm enc}/r^2}\notag\\
\sim&2 \left(\frac{\dot{E}}{10^{43}{\rm erg\, s}^{-1}}\right) \left(\frac{\kappa}{10^{29}{\rm cm}^2\,{\rm s}^{-1}}\right)^{-1} \left(\frac{r}{10\,{\rm kpc}}\right)^{-1}\notag\\ 
&\left(\frac{v_c}{500 {\rm km\,s}^-1}\right)^{-2} \left(\frac{n}{0.01 {\rm cm}^{-3}}\right)^{-1}.
\end{align}
This is consistent with our result that when the CR energy input reaches $\sim10^{43}{\rm erg\,s}^{-1}$, the CR pressure gradient starts to surpass the gravitational force in the core region and the core density and cooling rate start to be suppressed. 

We emphasize that, as shown in \fref{fig:e_compare}, the heating from streaming loss does exceed the cooling rate in the core region in  `CR-BH-43' (the most stably quenched CR injection case), which is consistent with the previous studies \citep[e.g.,][]{2017ApJ...844...13R}. However, we argue that the quenching is majorly caused by CR pressure lowering the gas density and therefore also the cooling rate instead of CR heating overcoming cooling because  a) the cooling rate of `CR-BH-43' is much lower than the  `Default' run; b) the heating from stellar feedback can be at least comparable to the CR heating in the core region; and c) the black hole thermal heating run with exactly the same energy input does not quench the galaxy.  

Given that we are not directly balancing the cooling rate by CR heating, the total CR energy in the halo does not need to be excessively high. The estimated $>$ GeV gamma-ray luminosity of `CR-BH-43' (from hadronic loss) is $L_{\gamma}\sim 10^{41}{\rm erg\,s}^{-1}$, which is lower than the observational upper bounds \citep{2016ApJ...819..149A,2018arXiv181202179W}. Besides, the estimated $\sim$ GHz radio luminosity of `CR-BH-43' from the secondary CR electrons (from CR protons), which contributes as part of the overall radio luminosity, is $L_{\rm radio}\sim10^{39}{\rm erg\,s}^{-1}$, again  within the observational constraint from the radio flux \citep[e.g.,][]{2014ApJ...781....9G,2016arXiv160300368B}.\footnote{We assume that all the secondary CR electrons decay via synchrotron emission.}

\section{Conclusions}\label{sec:conclusions}

In this paper, we have attempted a systematic exploration of different qualitative physical mechanisms by which energy can be injected into massive halos to quench galaxies and suppress cooling flows. We specifically considered models with radial momentum injection (e.g., ``wind'' or ``radiation pressure'' or ``isotropic kinetic'' models), thermal heating (e.g.\ ``shocked wind'' or ``isotropic sound wave'' or ``photo/Compton-heating'' or ``blastwave'' models), turbulent ``stirring'' (e.g.\ ``convective/buoyant bubble'' or ``precessing jet'' or ``jet/bubble instability-driven'' or ``subhalo/merger/satellite wind-driven'' models), and cosmic ray injection (e.g.\ CRs from compact or extended radio jets/lobes, shocked disk winds, or inflated bubbles). We vary the associated energetics and/or momentum fluxes, spatial coupling/driving scales, and halo mass scale from $\sim 10^{12}-10^{14}\,M_{\odot}$. These were studied in fully-global but non-cosmological simulations including radiative heating and cooling, self-gravity, star formation, and stellar feedback from supernovae, stellar mass-loss, and radiation, enabling a truly ``live'' response of star formation and the multi-phase ISM to cooling flows; we used a hierarchical super-Lagrangian refinement scheme to reach $\sim 10^{4}\,{\rm M}_{\sun}$ mass resolution, much higher than many previous global studies.

Of the cases surveyed, only turbulent stirring within a radius of order the halo scale radius, or cosmic ray injection (with appropriate energetics) were able to maintain a stable, cool-core, low-SFR halo for extended periods of time, across all halo masses surveyed, without obviously violating observational constraints on halo gas properties or exceeding plausible energy budgets for low-luminosity AGN in massive galaxies.


\begin{itemize}

\item{Isotropic momentum injection with momentum flux lower than $\sim10^{36} {\rm g\,cm\,s}^{-2}\,(M_{\rm halo}/10^{14}\,M_{\sun})^{1/3}$ has little effect on cooling flows or star formation, while larger momentum fluxes simply generate an ``explosion'' that evacuates gas from the halo core, drives strong shocks in the outer halo, generates steep negative temperature gradients out to $>100\,$kpc, and heats gas to enormous temperatures (all in conflict with observations).}

\item{Thermal heating, if concentrated in the halo core, similarly transitions sharply from doing nothing when the input is below cooling rates, to generating an explosive Sedov-Taylor blastwave when the input exceeds cooling rates (again, in conflict with observations). Thermal heating extended over too large a radius ``wastes'' all its energy at very large radii and does little in the core. It is possible to fine-tune thermal heating (by setting energy input equal to cooling rates, and the coupling scale equal to the cooling radius), but this (a) requires thermal heating rates $\gtrsim 10^{44}\,{\rm erg\,s^{-1}}$ in $\gtrsim 10^{14}\,M_{\sun}$ halos (corresponding to bright quasars if the heating efficiency is $\sim 1\%$), (b) still generates mild negative temperature gradients to $\sim 100\,$kpc, and (c) fails in less massive halos $\lesssim 10^{12.5}\,M_{\sun}$ where virial temperatures are lower.}

\item{Cosmic rays can suppress cooling and SFRs by supporting non-thermal pressure gradients which are comparable to or exceed gravity in the core, with modest energetics in an order-of-magnitude range around $\dot{E}_{\rm cr} \sim 10^{43}\,{\rm erg\,s^{-1}}\,(M_{\rm halo}/10^{14}\,M_{\sun})$. CR ``heating'' (via streaming or collisional terms) is negligible as modeled here in the interesting regime. For reasonable diffusivities the injection scale/kernel also does not matter sensitively since CRs form an equilibrium diffusion profile, unless the injection scale is very large $\gtrsim 30-100\,$kpc. The central few kpc tend to be ``hot'' because they are eventually depleted of all dense gas, but the larger-scale density/temperature/entropy structure of the cool-core halo can be stably maintained for extended periods of time, despite suppressed SFRs and actual cooling flow rates onto the galaxy.}

\item{Turbulent stirring can also suppress star formation, through a combination of suppressing the core gas density (by providing non-thermal pressure and ``lofting'' parcels up the potential where they buoyantly expand), and mixing cold and dense gas into the hot halo (providing ``bulk conduction''), with even lower energetics in $\dot{E}_{\rm turb} \sim 10^{41-42}\,{\rm erg\,s^{-1}}\,(M_{\rm halo}/10^{14}\,M_{\odot})$ or (equivalently) momentum flux $\dot{P}_{\rm turb} \sim 10^{34}\,{\rm g\,cm\,s^{-2}}\,(M_{\rm halo}/10^{14}\,M_{\odot})$ within a radius of order the halo scale radius ($\lesssim 100\,$kpc). Towards the low end of this range, halos maintain cool-core features, while towards the high end, they evolve from cool-core to non-cool-core. Strong stirring at $r\gtrsim 100$\,kpc tends to remove significant gas from the halo and suppresses the X-ray luminosity below observations; stirring confined only to $\lesssim 10-20$\,kpc acts more like galactic fountains and fails to efficiently suppress cooling. Turbulent ``heating'' (via compression or shocks or viscosity) is never dominant.}

\item{If injection transforms a halo into a non-cool-core, then if the core density is suppressed to an extent that the energy input from old stellar populations (SNe Ia and AGB mass-loss) exceeds cooling rates, the halo never ``rejuvenates'' even if the feedback injection shuts off.}

\end{itemize}

In summary, our study supports the idea that quenching -- at least of observed $z\sim0$ massive halos -- is not dominated by single violent or ``explosive'' events, but by lowering densities and suppressing cooling via mechanisms that involve relatively mild energetics and non-thermal pressure. Turbulence and cosmic rays represent  promising avenues to this, either of which has the potential to quench the models surveyed here without obviously contradicting basic observational constraints. Both operate very efficiently, with required energetics comparable to those expected in jets of low-luminosity AGN. 

We emphasize that we are not saying it is impossible to devise models of feedback using a combination of thermal and radial mechanical energy input which produce quenching and plausible massive halo properties (in fact, we explore a couple such models here). However, consistent with most previous studies, we find that these classes of models (a) require fine-tuning, in energetics and coupling scale as a function of halo mass, (b) generally require optimistically high energetics (at least order-of-magnitude larger than the CR models, and two orders of magnitude higher than the turbulent models favored here), and (c) may still have difficulty reproducing more subtle observational properties (e.g.\ distributions of temperature profile slopes). 

We should also emphasize that we are not implying that AGN feedback is represented by any one of these mechanisms (especially as we model them). Real feedback is a mix of many different processes operating at once, often simultaneously on very different scales (e.g.\ radiation and accretion-disk winds and jets may be coupling to the gas all on different spatial scales). Our goal was simply to focus on an (intentionally) highly-idealized model of each form of injection, to understand the constraints and different qualitative behaviors of different types of energy injection. This paper was a follow-up to \paperone, where we also surveyed a large number of simulations to emphasize that {\em something} beyond the ``default'' physics of cooling, self-gravity and gravitational stability, magnetic fields, conduction, viscosity, star formation, and feedback from stars (radiative and supernovae and stellar mass-loss), was required to resolve the cooling flow problem. Here we identify plausible {\em classes} of physical candidates for that ``something'' (e.g., enhanced turbulence and CR from AGN). In our next study, we intend to model these classes more realistically: for example, explicitly modeling a narrow jet which simultaneously carries kinetic luminosity and cosmic rays. This raises a host of questions we have (again, intentionally) not tried to address here: for example, what happens if the injection is highly anisotropic? And can turbulence actually be driven by {\em physical} processes originating from an AGN? And what is the ratio of energy in radial momentum flux, thermal heating, cosmic ray injection, and turbulent stirring which comes from e.g.\ nuclear winds vs.\ compact jets vs.\ ``bubbles''? These and many more questions remain open and critical for progress in this field.

\acknowledgments
We thank Eliot Quataert for conversations and collaboration.  Support for PFH was provided by an Alfred P.~Sloan Research Fellowship, NASA ATP Grant NNX14AH35G, and NSF Collaborative Research Grant \#1411920 and CAREER grant \#1455342. The Flatiron Institute is supported by the Simons Foundation. CAFG was supported by NSF through grants AST-1517491, AST-1715216, and CAREER award AST-1652522, by NASA through grant 17-ATP17-0067, by CXO through grant TM7-18007, and by a Cottrell Scholar Award from the Research Corporation for Science Advancement.
DK was supported by NSF grant AST-1715101 and the Cottrell Scholar Award from the Research Corporation for Science Advancement. TKC was supported by NSF grant AST-1412153.
V.H.R. acknowledges support from UC-MEXUS and CONACyT through the postdoctoral fellowship.
Numerical calculations were run on the Caltech compute cluster ``Wheeler,'' allocations from XSEDE TG-AST130039 and PRAC NSF.1713353 supported by the NSF, and NASA HEC SMD-16-7592. \\

\bibliographystyle{mnras}
\bibliography{mybibs}

\appendix\label{appendix}
\normalsize

\section{Effects of Magnetic Fields, Conduction, and Viscosity on Turbulent ``Stirring'' Models}

\begin{figure}
\centering
\includegraphics[width=8.5cm]{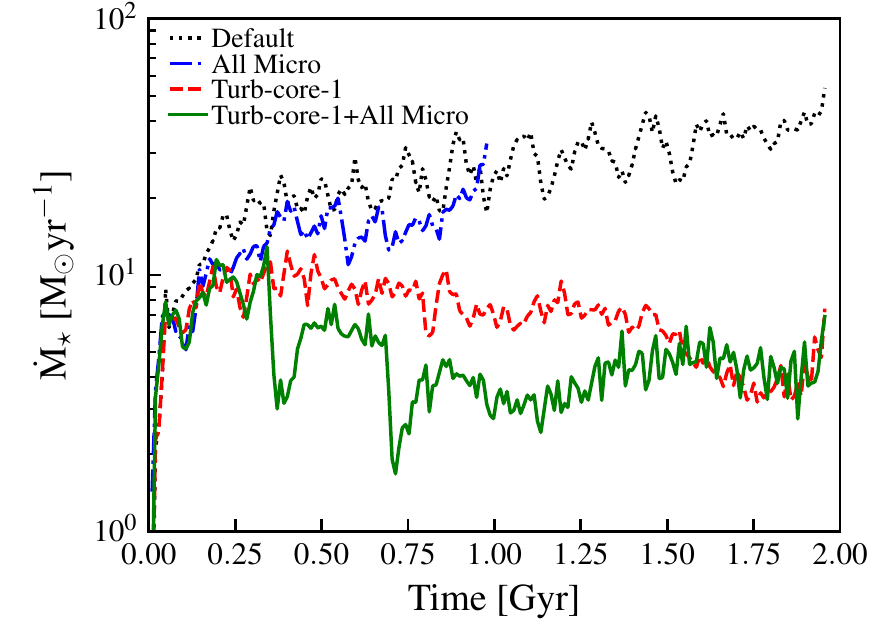}
\caption{SFR (as Fig.~\ref{fig:sfr}) in our `Default' and `Turb-core-1' {\bf m14} runs, comparing runs which treat the gas as pure-hydrodynamic, to runs which include magnetic fields and fully-anisotropic Spitzer-Braginski conduction and viscosity following \citet{2016arXiv160705274S} (``All Micro''). Consistent with our study in \paperone, these additional microphysics (mostly conduction) suppress the SFRs by a factor $\sim 2$, but do not qualitatively change any of our conclusions. 
}
\label{fig:sfrmcvt}
\end{figure}

Given that turbulent stirring can (a) amplify magnetic fields, (b) be damped by viscosity from the hot gas, and (c) acts to mix hot and cold gas in a manner similar to physical conductivity, it is reasonable to ask what the impact of including or excluding explicit treatment of magnetic fields and physical (anisotropic) Braginskii conduction and viscosity in the hot gas might be. We explored these physics in \paperone\ in detail so only briefly note their effects here. \fref{fig:sfrmcvt} shows the SFRs of the `Turb-core-1' run with and without explicit inclusion of these fluid microphysics in the simulations. Magnetic fields and conduction mildly suppress the SFR at the beginning of the `Turb-core-1' run, and suppress the core baryonic mass by a factor of $\sim 2$, which is roughly  consistent with their effect on the `Default' run, but the systematic effects are small and get {\em smaller} as time goes on and the systems become more steady-state. Because viscosity and conduction are strongly temperature-dependent, their effects are even weaker in the smaller halo masses. Accordingly, the treatment of these physics does not substantially alter our conclusions.


\label{lastpage}

\end{document}